\documentclass[
 aip,
 jmp,%
 amsmath,amssymb,
 preprint,%
]{revtex4-1}

\newcommand{\be}{\begin{equation}}
\newcommand{\ee}{\end{equation}}

\newcommand{\ba}{\begin{array}}
\newcommand{\ea}{\end{array}}
\newcommand{\bea}{\begin{eqnarray}}
\newcommand{\eea}{\end{eqnarray}}
\newcommand{\nn}{\nonumber \\}
\newtheorem{Theorem}{Theorem}

\usepackage{dcolumn}
\usepackage{bm}

\usepackage{amsfonts}    
\usepackage{amssymb}
\usepackage{amsmath}
\usepackage{latexsym}
\usepackage{eepic}
\usepackage{graphicx}
\usepackage[usenames,dvipsnames]{color}
\usepackage{color}

\newlength\tindent
\newtheorem{conjecture}{Conjecture}

\begin{document}


\title[{\footnotesize}]{New infinite families of $N$th-order superintegrable systems separating in Cartesian coordinates}

\author{A. M. Escobar-Ruiz}
\email{admau@xanum.uam.mx}
\affiliation{
Departamento de F\'isica,
Universidad Aut\'onoma Metropolitana-Iztapalapa, San Rafael Atlixco 186,
M\'exico, CDMX, 09340 M\'exico}

\author{R. Linares}
\email{lirr@xanum.uam.mx}
\affiliation{
Departamento de F\'isica,
Universidad Aut\'onoma Metropolitana-Iztapalapa, San Rafael Atlixco 186,
M\'exico, CDMX, 09340 M\'exico}

\author{P. Winternitz}
\email{wintern@crm.umontreal.ca}
\affiliation{
Centre de recherches math\'ematiques
and D\'epartement de math\'ematiques \\
et de statistique, Universit\'e de Montr\'eal,  C.P. 6128,
succ. Centre-ville,\\ Montr\'eal (QC) H3C 3J7, Canada}

\date{\today}

\begin{abstract}
A study is presented of superintegrable quantum systems in two-dimensional Euclidean space $E_2$ allowing the separation of variables in Cartesian coordinates. In addition to the Hamiltonian $H$ and the second order integral of motion $X$, responsible for the separation of variables, they allow a third integral that is a polynomial of order $N\, (N\geq3)$ in the components $p_1, p_2$ of the linear momentum. We focus on doubly exotic potentials, i.e. potentials $V(x, y) = V_1(x) + V_2(y)$ where neither $V_1(x)$ nor $V_2(y)$ satisfy any linear
ordinary differential equation. We present two new infinite families of superintegrable systems in $E_2$ with integrals of order $N$ for which $V_1(x)$ and $V_2(y)$ are given by the solution of a nonlinear ODE that passes the Painlev\'e test. This was verified for $3\leq N \leq 10$. We conjecture that this will hold for any doubly exotic potential and for all $N$, and that moreover the potentials will always actually have the Painlev\'e property.
\end{abstract}

\maketitle

\section{INTRODUCTION}
\label{intro}

In this article we concentrate on systems that are separable in Cartesian coordinates
and possess an additional integral of arbitrary order $N>2$. The systems are
second-order integrable because in addition to the Hamiltonian
\begin{eqnarray}
{H}\  = \ \frac{1}{2}\bigg(\,p_1^2 \ + \ p_2^2\,\bigg) \ + \ V(x,y)\ ,
\qquad
V(x,y)\,  =\,  V_1(x) \, + \, V_2(y)\ ,
\label{Hcart}
\end{eqnarray}
they allow, for arbitrary potential functions $V_1(x)$ and $V_2(y)$, a second-order integral
\begin{eqnarray}
X\ = \ \frac{1}{2}\bigg(\,p_1^2 \ - \ p_2^2\,\bigg) \ + \ V_1(x) \ - \ V_2(y)  \ .
\label{X}
\end{eqnarray}
Evidently, the Hamiltonian (\ref{Hcart}) is ${\mathcal{S}}_2$ invariant under the permutation $x \Leftrightarrow y $ while $X$ is anti-invariant. The existence of an additional integral $Y$, a polynomial of order $N$ in the momentum components, makes the system $N$th-order superintegrable
(more integrals of motion than degrees of freedom). In classical mechanics $H$, $X$ and $Y$ are well-defined functions on phase space and are functionally independent. In quantum mechanics they are assumed to be polynomials, or convergent series\cite{Eichler, Magnus} in the enveloping algebra of the Heisenberg algebra in $E_2$, {\it i.e.}, the Lie algebra with basis $\{x, y, p_1, p_2, 1\}$. The operators $H$, $X$, $Y$ are assumed to be polynomially independent, that is no nontrivial Jordan polynomial in the quantities $(H, X, Y)$ is equal to zero\cite{Burchnall,Chalykh,Hietarinta1998, Weigert}. We will use the usual vector fields $(p_{1} = - i \,\hbar\, \partial_x,\,  p_{2} =- i \,\hbar\, \partial_y)$ in quantum mechanics. In classical mechanics, $p_1$ and $p_2$ are the canonical momenta components conjugate to the coordinates $x$ and $y$, respectively.

For recent reviews of classical and quantum superintegrable systems see the articles\cite{MillerPostWinternitz:2013, Millerebook}. In particular, for the intimate relationship between superintegrability and exact solvability in quantum mechanics see the papers\cite{TempTW, Tremblay, Fordy1,Fordy2}. Theoretical studies of integrable and superintegrable systems have been extended to Riemannian and pseudo-Riemannian spaces of arbitrary dimensions and with integrals of arbitrary order as well\cite{Benenti,Chanu,Gonera,Iliev, Kress, Shmavonyan}. For recent applications of superintegrable systems in such diverse fields as optics, condensed-matter physics and the theory of orthogonal polynomials see the articles\cite{Atakishiyev, Genest, Nikitin2, Pogosyan, Tyc}.

An interesting phenomenon was observed in several works\cite{gW,gWW, TPW2010} when studying third order ($N=3$) superintegrable quantum systems in $E_2$. Namely, when the potential allows a second order integral (that leads to separation of variables in either Cartesian or polar coordinates) and an additional one of (at least) third order then \emph{exotic potentials} arise, (see \cite{AW, gW,TPW2010, AMJVPW2015, EWY}). These are potentials that do not satisfy any linear ordinary differential equation (ODE) but only nonlinear ones. It turned out that all the ODEs obtained in the quantum case for these potentials pass the Painlev\'e test, a necessary condition for the Painlev\'e property. In fact, most of the ODEs obtained in the quantum case were shown to have the Painlev\'e property. The ODEs were explicitly integrated in terms of the 6 original Painlev\'e transcendents\cite{Ince:ode, Painleve:Pain, Gambier:Pain}. This means that the general solution (in the case of the present article the potential) of these equations has no movable critical singularities (see Refs.\cite{Ince:ode, Painleve:Pain, Gambier:Pain, Conte:R99, Conte}). It can hence be expanded into a Laurent series with a finite number of negative powers.

Another important fact is that the exotic quantum potentials are proportional to $\hbar^2$. The appearance of $\hbar-$dependent potentials is not as rare (or unphysical) as one may think. In a series of papers of one of the authors (see \cite{MTE:2018, MTE:20182, TME3B} and references therein), the quantum $n$-body problem in $d-$dimensional Euclidean space with interaction depending only on mutual (relative) distances was considered. It was shown that for the ground state a reduced Hamiltonian can be constructed. This Hamiltonian describes a $\frac{n(n-1)}{2}$-dimensional quantum particle moving in a curved space with an additional $d$-dependent singular effective potential $V_{\rm eff}$. Remarkably, this potential $V_{\rm eff}$ is also proportional to $\hbar^2$.

The aim of this article is to establish some general properties of
superintegrable systems separating in Cartesian coordinates and allowing
a higher $N$th order integral. The properties were observed for specific
choices of $N$. The method and results presented here can be easily extended to all
$N\geq 3$.

Among the results observed for $3\leq N \leq 5$ we mention:
\begin{enumerate}
\item Superintegrable Hamiltonians in classical and quantum
mechanics can differ \cite{Hietarinta1998, HietarintaGrammaticos}. Terms depending on $\hbar$ can appear
in the quantum case. The classical limit $\hbar \rightarrow 0$ can
be singular and must be taken in the determining equations, not in
the solutions. This is true for any Hamiltonian in $E_2$ with an integral
of order $N$, independently of the separation of variables.
\item Three types of potentials and superintegrable systems occur. We call them doubly exotic, singly exotic and standard. For standard systems the potential functions $V_1(x)$ and $V_1(y)$ are both solutions of linear ODEs obtained from a linear compatibility condition for the nonlinear determining equations that must be satisfied for $Y$ to be an integral of motion. For doubly exotic potential the linear ODEs are satisfied trivially for both potential functions, i.e. the integral $Y$ is such that all coefficients in the linear compatibility condition vanish. The functions $V_1(x)$ and $V_1(y)$ then satisfy nonlinear ODEs. In quantum mechanics, these nonlinear equations pass the Painlev\'e
test \cite{ARS, Conte}, in the classical case they do not. In general, the quantum doubly exotic potentials were then expressed in terms of elliptic functions, or known (second order) Painlev\'e transcendents (i.e. the solutions of the Painlev\'e equations  \cite[page 345]{Ince:ode}). However, starting from $N=5$ it was found\cite{AW,GungorKNN:2014} that certain potentials cannot be expressed in terms of the six Painlev\'e transcendents. It is conjectured that they define new (higher order) Painlev\'e transcendents.
\item The integrals of motion $H$, $X$ and $Y$ satisfy $[H, X] = [H, Y] = 0$, $[X, Y] = C \neq 0$,
where $[\cdot, \cdot]$ denotes the Lie bracket in quantum mechanics
and the Poisson bracket in the classical case. Further commutations like
$[X,C]$, $[Y, C], \ldots,$ in general yield an infinite dimensional Lie algebra,
exceptionally a finite dimensional algebra \cite{Bargmann, Fock, Jauch},
or a Kac-Moody algebra \cite{DSD}. It
is however more fruitful to view this algebra as a finite dimensional polynomial
Lie or Poisson algebra \cite{Daskol, Milleretal, IanM, Chen, Latini}. In many cases
it turns out that the commutators $[X,C] = D_1$, $[Y, C] = D_2$
are polynomials in $X$, $Y$ and $H$ with constant coefficients.

\item A simple and crucial observation is that if a potential $V(x,y)$ as in (\ref{Hcart})
allows an integral $Y$ of order $M$ then the same potential will show up
again for infinitely many values of $N \geq M$. This is because all
powers $X^a \,Y^b\,H^c$ and all commutators $[X^a, Y^b]$
(Lie or Poisson, respectively) are also integrals of motion. Only $3$ of them
(including the Hamiltonian) can be functionally (or polynomially) independent.
What is of interest is to determine the lowest order of $N$ for each
superintegrable potential. As an example we recall that Drach in his
pioneering article \cite{Drach1} found $10$
classical complex integrable potentials with a third-order integral (in $E_2( \mathbb {C})$).
Much later it was shown \cite{Ranada, Tsiganov:2000} that $7$ of the $10$ systems were
reducible. Indeed, these $7$ were actually second-order superintegrable and the
third-order integral is a commutator of two second-order ones.
\end{enumerate}
The above features of quantum superintegrable systems in two-dimensional Euclidean spaces are shared by all known superintegrable systems. In particular, a new \emph{Painlev\'e conjecture}\cite{MPPC} states that if the potentials satisfy identically a certain linear compatibility condition for the existence of an integral of motion of order $N\geq 3$ and also allow the separation of variables in Cartesian (or polar) coordinates, then they will be solutions of ordinary differential equations that pass the Painlev\'e test. All linear equations have the Painlev\'e property by default, they have no movable singularities at all. Exotic potentials, on the other hand, are solutions of genuinely nonlinear ODEs that pass the Painlev\'e test and in most cases have been shown to have the Painlev\'e property. This represents a surprising connection of higher order superintegrability in the quantum case with soliton theory of infinite-dimensional integrable nonlinear systems.

The motivation of this article is twofold, first to shed light on the classification of general higher order two-dimensional superintegrable systems and, secondly, to describe a systematic method to calculate them. A recent different approach based on operator algebras is analyzed in the articles\cite{GungorKNN:2014, MSW2}, see also the paper\cite{Nikitin} for the one-dimensional case.

A general observation is that standard superintegrable potentials are determined by linear ODEs arising from a linear compatibility condition (LCC). For doubly exotic potentials the LCC is satisfied trivially but new nonlinear compatibility conditions (NLCC) arise that determine the potentials.

The structure of the article is the following. In Section \ref{integralN}, we first review the determining equations governing the existence (and the form) of a general $N$th-order polynomial integral $Y_N$ for a separable potential as in eq. (\ref{Hcart}). In Section \ref{coefj0} we show that the highest order terms in the integral $Y_N$ lie in the $N$th layer of the enveloping algebra
of the Euclidean Lie algebra  $e(2)$. The next to leading terms are studied in Section \ref{LCC NLCC} where we show that a LCC that is a linear PDE for any superintegrable potential reduces to two uncoupled linear ODEs for each of the potential functions $V_1(x)$ and $V_2(y)$ in the separable case (\ref{Hcart}). In Section \ref{coefj4} we introduce a “well” of determining equations and NLCC. The potentials are determined from the compatibility conditions. Once the potentials are known, the remaining determining equations become linear and can be solved.
In Section \ref{families} the LCC is used to classify superintegrable systems into 3 classes:
doubly exotic, singly exotic and standard. A general formula for $Y_N$ is given for the doubly exotic case. Two types of trivial integrals are
discussed in Section \ref{trivint}. The first is due to separability alone and the trivial integral $Y_N^{tr}$ is a polynomial in $H$ and $X$. The second type is a consequence of lower order superintegrability. Both types should be discarded. In Section \ref{partcases} we introduce two infinite families of doubly exotic nontrivial $N$th-order systems. Section \ref{DoublyExoticlow} is devoted to low order examples with $N= 3,4,\dots,10$. For conclusions and future outlook see Section \ref{Conclusions}.

\section{General $N$th-order integral for potentials allowing separation of variables in Cartesian coordinates}
\label{integralN}
\subsection{General form}

In the quantum case, we can write the most general $N$th-order Hermitian polynomial operator $Y_N$ in the form
\begin{equation}\label{YNQSd}
Y_N \ =  \ \frac{1}{2}\sum_{\ell=0}^{[\frac{N}{2}]}\sum_{j=0}^{N-2\ell}\,\{\,f_{j,2\ell}\,,\, p_1^j\ p_2^{N-j-2\ell}\,\} \ ,
\end{equation}
(cf. eq. (6) in Ref.\cite{PostWinternitz:2015}), where $N \in \mathbb{Z}^+$, $p_{1} = - i \,\hbar\, \partial_x,\,  p_{2} =- i \,\hbar\, \partial_y$, $[a]$ indicates the integer part of $a$, $\{\,,\,\}$ denotes an anticommutator and ${f}_{j,2\ell}={f}_{j,2\ell}(x,y,V)$ are real functions that, in general, depend on the variables $x$ and $y$ and the potential $V$. These functions as well as the potential $V$ will be determined by vanishing the commutator with the Hamiltonian $[H,\,{Y}_N]=0$.

Equivalently, the integral $Y_N$ (\ref{YNQSd}) can be written as follows
\begin{equation}\label{Yq}
Y_N \ = \ W_N  \ + \ \frac{1}{2}\sum_{\ell=1}^{[\frac{N}{2}]}\sum_{j=0}^{N-2\ell}\,\{\,{\tilde f}_{j,2\ell}\,,\, p_1^j\ p_2^{N-j-2\ell}\,\} \qquad ,
\end{equation}
where the $N$th-order terms are collected in $W_N$
\begin{equation}\label{YNq}
W_{N}  \ = \  \frac{1}{2}\sum_{0\leq m+n\leq N}^{}\ A_{N-m-n,m,n}\ \{\,L_z^{N-m-n}\,, \ p_1^m\,p_2^n\,\}  \ .
\end{equation}
In the above formulas $L_z=x\,p_2-y\,p_1$ is the angular momentum operator and $A_{N-m-n,m,n}$ are $\frac{(N+1)(N+2)}{2}$ arbitrary real constants with at least one of them different from zero so that the integral $Y_N$ (\ref{YNQSd}) is of order $N$. The functions ${\tilde f}_{j,2\ell}$ in (\ref{Yq}) depend on the variables $x$, $y$ and the potential $V$ as well. This term $W_N$ is fundamental since it determines the existence of the $N$th-order integral $Y_N$ (\ref{YNQSd}). We will see that the potential $V$ (\ref{Hcart}) obeys only non-linear ODEs depending on which parameters $A_{N-m-n,m,n}$ in (\ref{YNq}) are present or not.

In $Y_N$ (\ref{YNQSd}), by putting $N=2$, $A_{0,2,0}=-A_{0,0,2}=\frac{1}{2}$, $A_{2,0,0}=A_{1,1,0}=A_{1,0,1}=A_{0,1,1} =0$ and ${\tilde f}_{0,2}=V_1(x)-V_2(y)$  we arrive to second order integral $X$ (\ref{X}), namely $Y_2=X$.

The second order \emph{integrable} Hamiltonian (\ref{Hcart}) becomes $N$th-order ($N>1$) \emph{superintegrable} if it commutes with the operator $Y_N$.

\subsection{The determining equations}
For an arbitrary $N$th-order polynomial differential operator
\[
{\cal Y}_N \  = \ \sum_{k+l=0}^{N}\{\,g_{k,l}(x,y)\,,\, p_1^k\ p_2^{l}\,\} \ ,
\]
the commutator with the Hamiltonian $[H,\,{\cal Y}_N]$ is a differential operator of order $(N+1)$, {\it i.e.} we have
\begin{equation}\label{HYC}
  [H,\,{\cal Y}_N] \ = \  \sum^{N+1}_{k+l=0} \   Z_{k,l}\ \frac{\partial^{k+l}}{\partial x^k\,\partial y^l} \ ,
\end{equation}
where the coefficients $Z_{k,l}=Z_{k,l}(x,\,y;\,g_{k,l},\,V)$, in front of the partial derivatives, depend on $x,y$, $g_{k,l}$ and the potential $V$. We require
\begin{equation}\label{Zxy}
Z_{k,l}(x,\,y;\,g_{k,l},\,V)\ = \ 0  \qquad \ \text{for all $k$ and $l$}\ ,
\end{equation}
($[H,\,{\cal Y}_N ]=0$) and obtain the \emph{determining equations}. For an $N$th-order integral of the form (\ref{YNQSd}), and an arbitrary potential $V$, these equations were derived in Ref.\cite{PostWinternitz:2015}.

In the case of a separable potential $V(x,\,y) \, = \, V_1(x) \, + \, V_2(y)$, the determining equations in Ref.\cite{PostWinternitz:2015} reduce to

{
\be\label{quant deteq} 0 \ = \   {M}_{j,2\ell}    \  \equiv \  Z_{j,\,N-2\,\ell-j+1}  \  ,\ee
}
where
\bea  \label{Mj2l} {M}_{j,2\ell}& \ \equiv \  & 2\left( \partial_x{ f}_{j-1,2\ell} \ + \ \partial_y{f}_{j,2\ell}\right) \\
 &&-\left(2(j+1){f}_{j+1, 2\ell-2} V'_1\ + \ 2(N-2\ell+2-j){f}_{j, 2\ell-2} V'_2 \ + \ \hbar^2 {Q}_{j,2\ell}\right), \nonumber \eea
here ${ Q}_{j,2\ell}$ is a quantum correction term given by
\bea \label{Quantcorrection} {Q}_{j,2\ell}  && \  \equiv \ \left(2\partial_x{\phi}_{j-1,2\ell}\ + \ 2\partial_y{\phi}_{j,2\ell} \ + \ \partial_x^2{ \phi}_{j,2\ell-1}\ + \ \partial_y^2{\phi}_{j,2\ell-1}\right)\\
 &&- \ 2\sum_{n=0}^{\ell-2}(-\hbar^2)^n\bigg[\ \binom{N-2\ell+2n+4-j}{ 2n+3}\,V_2^{(2n+3)}\,{f}_{j,2\ell-2n-4} \ + \ \nn
 && \binom{j+2n+3}{2n+3}\,V_1^{(2n+3)}\,{f}_{j+2n+3,2\ell-2n-4}\ \bigg]\nn
 &&-\ 2\sum_{n=1}^{2\ell-1}(-\hbar^2)^{\lfloor(n-1)/2\rfloor}\,\bigg[ \ \binom{ N-2\ell+n+1-j}{ n}\,V^{(n)}_2\,{\phi}_{j,2\ell-n-1} \ +\ \nn
 && \binom{j+n}{n}\,V^{(n)}_1\,{\phi}_{j+n,2\ell-n-1}\ \bigg]\ , \nonumber\eea
$V_1^{(s)}\equiv \frac{d^s}{dx^s}V_1$, $V_2^{(q)}\equiv \frac{d^q}{dy^q}V_2$. The functions ${\phi}_{j,k}$ are defined for $k>0$ as
\be  {\phi}_{j,2\ell-\epsilon} =\sum_{b=1}^{\ell}\sum_{a=0}^{2b-\epsilon}\frac{(-\hbar^2)^{b-1}}{2}\binom{j+a}{a}\binom{ N-2\ell+2b-j-a}{2b-\epsilon-a}\partial_x^a\partial_y^{2b-\epsilon-a}{f}_{j+a, 2\ell-2b} \label{phieven} \ ,\ee
with $\epsilon=0,1.$ The real functions ${f}_{j, \ell} \equiv 0$ identically for $\ell<0$ and $j < 0$ as well as for $j > N - 2\ell$.
\begin{itemize}
  \item All other determining equations ${M}_{j,\,2\ell+1}=0$ can be written as differential consequences of the set of equations (\ref{quant deteq}).
  \item The functions $f_{j,0}$, those with $\ell=0$, do not depend on the potential and can be solved explicitly. They are presented in the next Section.
  \item From (\ref{quant deteq}) with $\ell=1$, we obtain a \emph{linear compatibility condition} for the potential $V$ only. These equations also define the functions ${f}_{j, 2}$ in terms of $x$,$y$ and the potential $V$.
  \item Starting from $\ell=2$, the equations (\ref{quant deteq}) will lead to \emph{nonlinear compatibility conditions} for the potential $V$ only.
\end{itemize}

The number of determining equations (\ref{quant deteq}) is equal to
\be\label{numdet} \sum_{\ell=0}^{[\frac{N+1}{2}]}(N-2\ell +2)=\left\{ \ba{ll} \frac14(N+3)^2 & \ \mbox{N odd}\\
\frac14(N+2)(N+4) &  \ \mbox{ N even.}\ea \right.\ee
For a given potential $V(x,y)$ the determining equations are linear first-order partial differential equations for the unknowns $ f_{j,2\ell}(x,y)$. The number of unknowns is
\be \label{numcof} \sum_{\ell=0}^{[\frac{N+1}{2}]}(N-2\ell +1)=\left\{ \ba{ll}\frac{(N+1)(N+3)}4 \qquad \ & \mbox{N odd}\\
\frac14(N+2)^2 \qquad & \mbox{ N even \ .}\ea \right.\ee
The system is overdetermined and subject to compatibility conditions.
If the potential $V(x,y)$ is not a priori known, then the system (\ref{quant deteq}) becomes nonlinear
and $V(x,y)$ must be determined from the compatibility conditions.

The determining equations (\ref{quant deteq})-(\ref{Quantcorrection}) are written for $Y_N$ in the form (\ref{YNQSd}), that is, for the functions $f_{j,2\ell}$. The equivalent ones for the functions $\tilde f_{j,2\ell}$ in (\ref{Yq}) can be obtained in a similar way, see eqs. $(39)$-$(42)$ in Ref\cite{PostWinternitz:2015}).

In the classical case, the determining equations can be obtained from those of the quantum case by taking the appropriate limit $\hbar\rightarrow 0$\,.

\section{The coefficients $f_{j,0}$ and the enveloping algebra of the Euclidean Lie algebra $e(2)$}
\label{coefj0}
The functions $f_{j,0}$, which define the (leading) $N$th-order terms in $Y_N$ (\ref{YNQSd}), are given by the determining equations (\ref{quant deteq}) with $\ell=0$, $M_{j,0}=0$. These equations correspond to the vanishing of all the coefficients, in the commutator $[H,\,Y_N]$,
multiplying the partial derivatives of order $N+1$. They do not depend on the potential and can be solved explicitly. The solutions are given by\cite{PostWinternitz:2015}
\be \label{fj0 Nthorder}  f_{j,0}\ = \ \sum_{n=0}^{N-j} \sum_{m=0}^{j}\binom{ N-n-m}{j-m}A_{N-n-m,m,n}x^{N-j-n}(-y)^{j-m},\ee
which implies that the leading part (\ref{YNq}) of the integral $Y_N$ is a polynomial of order $N$ in
the enveloping algebra of the Euclidean Lie algebra $e(2)$ with basis $\{p_1, p_2 , L_z\}$. Thus, for any $N$ the functions $f_{j,\ell}$ with $\ell=0$ are known in terms of the $\frac{(N+1)(N+2)}{2}$ constants $A_{N-n-m,m,n}$ figuring in (\ref{YNq}).

\section{The coefficients $f_{j,2}$ and the linear compatibility condition}\label{LCC NLCC}

The functions $f_{j,2}$, which define the next-to-leading terms in $Y_N$ (\ref{YNQSd}), are defined by (\ref{quant deteq}) with $\ell=1$. These equations correspond to the vanishing of all the coefficients in the commutator $[H,\, Y_N]$ multiplying the partial derivatives of order $N-1$. These equations do depend on the potential $V(x,y)$. Hence, they can not be solved in full generality. However, for arbitrary $N$ it has been shown \cite{Hietarinta1987,PostWinternitz:2015} that their compatibility condition implies that the potential $V$, independently of the separation of variables, must satisfy a linear compatibility condition (LCC), a PDE of order $N$. This LCC is a necessary but not sufficient condition for the existence of the $N$th order integral. For a separable potential $V(x,\,y) \, = \, V_1(x) \, + \, V_2(y)$, the LCC takes the form
\begin{equation}\label{LCC00}
\begin{aligned}
\sum_{j=0}^{N-1}{(-1)}^j\,
\bigg[ & \ (j+1)\,\mathcal{A}_{j+1,N}\,{(-1)}^{(j+1-m)}\,\bigg[\bigg(\frac{d}{dy}\bigg)^j {y}^{j+1-m}\bigg]\,Q_1^{(j)}(x)   \quad + \quad
\\ &  (N-j)\,\mathcal{A}_{j,N}\bigg[\bigg(\frac{d}{dx}\bigg)^{N-j-1} {x}^{N-j-n}\bigg]\,Q_2^{(j)}(y)   \ \bigg]  \ = \ 0   \ ,
\end{aligned}
\end{equation}
where
\[
\mathcal{A}_{j,\,N}  \ \equiv \ \sum_{n=0}^{N-j}\sum_{m=0}^{j}\binom{N-m-n}{j-m}\ A_{N-m-n,\,m,\,n} \qquad \ ,
\]
and
\[
Q_1^{(j)}(x) \ \equiv \ \bigg(\frac{d}{dx}\bigg)^{N-j-1} [\,x^{N-j-n-1}\,  V_1'(x)\,] \ ,
\]
depends only on $x$, $V_1(x)$ and its derivatives, while
\[
Q_2^{(j)}(y) \ \equiv  \  {(-1)}^{(j-m)}\,\bigg(\frac{d}{dy}\bigg)^{j}  [\,y^{j-m}\,  V_2'(y)\,] \ ,
\]
is a function of $y$, $V_2$ and its derivatives alone. In (\ref{LCC00}), the $y$-contributions to the coefficient of $Q_1^{(j)}(x)$, those that depend on the variable $y$, come from the terms with $m=0$ and $m=1$ only, while for the coefficient of $Q_2^{(j)}(y)$ its $x$-dependence is due to the terms with $n=0$ and $n=1$ alone. Therefore, for any $j$ the coefficients in front of $Q_1^{(j)}(x)$ and $Q_2(y)^{(j)}$ are at most linear in the variables $y$ and $x$, respectively.

Hence, differentiating (\ref{LCC00}) twice with respect to $x$, we eliminate the dependence on $V_2(y)$ completely. The resulting equation is a polynomial of degree one in $y$ with coefficients $ \tau_1$ and $\upsilon_1$, respectively, that depend on both $x$ and derivatives of $V_1(x)$ only,
\begin{equation}\label{QX}
\partial^2_x\,(\text{LCC}) \ = \ \tau_1(x) \ + \ y\,\upsilon_1(x)  \ = \ 0 \ ,
\end{equation}
here
\[
\tau_1(x) \ = \ \sum_{j=0}^{N-1}(j+1)! \sum_{n=0}^{N-j-1}\binom{N-1-n}{j}\ A_{N-1-n,\,1,\,n}\,\bigg(\frac{d}{dx}\bigg)^{N-j+1} [\,x^{N-j-n-1}\,  V_1'(x)\,] \ ,
\]
and
\[
\upsilon_1(x) \ = \ \sum_{j=0}^{N-1}(j+1)(j+1)! \,(-1)^{2j+1}\, \sum_{n=0}^{N-j-1}\binom{N-n}{j+1}\ A_{N-n,\,0,\,n}\,\bigg(\frac{d}{dx}\bigg)^{N-j+1} [\,x^{N-j-n-1}\,  V_1'(x)\,] \ .
\]
Each of these two coefficients must vanish and we obtain two linear ODEs of order $(N+2)$ for $V_1(x)$, namely $\tau_{1}=0$ and $\upsilon_1=0$. Similarly, from the equation
\begin{equation}\label{QY}
\partial^2_y\,(\text{LCC}) \ = \ \tau_2(y) \ + \ x\,\upsilon_2(y)  \ = \ 0 \ ,
\end{equation}
we obtain two linear ODEs of order $(N+2)$, $\tau_2=0$ and $\upsilon_2=0$, for $V_2(y)$. They can be obtained from (\ref{QX}) using the symmetry $x\leftrightarrow y$. We have arrived at the following theorem.

\begin{Theorem}
A necessary condition for a separable potential $V(x,y)= V_1(x)
+ V_2(y)$ in $E_2$ to allow a polynomial integral of order $N\geq 3$ is that the
potential functions $V_1(x)$ and $V_2(y)$ satisfy the ODEs (\ref{QX}) and (\ref{QY}).
\end{Theorem}

\textbf{Remark}: The important point of Theorem 1 is that the number of ODEs is
always two, independently of the value of $N$. This implies that each potential function $V_1(x)$ and $V_2(y)$ satisfies an
overdetermined system of two linear ODEs, namely $\tau_{1}(x)=\upsilon_1(x)=0$ and $\tau_{2}(y)=\upsilon_2(y)=0$, respectively. This in turn explains why superintegrable systems are rare.

Moreover, in addition to the LCC (\ref{LCC00}), \emph{there exist nonlinear compatibility conditions which are very instrumental in finding superintegrable potentials}. So far, they have not been studied in detail. We will pay special attention to these nonlinear compatibility conditions.

\section{The coefficients $f_{j,4}$ and the nonlinear compatibility conditions}
\label{coefj4}

In this section, assuming a general $N$th order integral of the form (\ref{YNQSd}), we will construct a nonlinear compatibility condition for the potential alone. This equation is obtained from the determining equations (\ref{quant deteq}) with $\ell=2$, i.e. those that also define the functions ${f}_{j,4}$ in (\ref{YNQSd}). They correspond to the vanishing of all the coefficients, in the commutator $ [H,\,{Y}_N]$, multiplying the partial derivatives of order $N-3$.

For arbitrary $N>3$, the set of determining equations with $\ell=2$ is generically given by
\begin{equation}\label{NLCCG}
  \partial_x{ f}_{j-1,4} \ + \ \partial_y{f}_{j,4}  \ = \  F_{j}  \qquad ,\ \qquad j=0,1,2\ldots,N-4 \ \ ,
\end{equation}
see eqs.(\ref{quant deteq})-(\ref{Mj2l}), where
\begin{equation}\label{FJg}
F_j \ = \ F_j(x,y,V,{f}_{j,2},A_{N-m-n,\,m,\,n},N) \ ,
\end{equation}
depend on the potential $V$ (and its derivatives), the functions ${f}_{j,2}$ and the parameters $A_{N-m-n,\,m,\,n}$ and $N$. In (\ref{NLCCG}), ${f}_{j,2}$ is identically $0$ for $j < 0$ or $j > N - 4$. In particular, $F_j$ (\ref{FJg}) contain products between the derivatives of $V$ times the functions ${f}_{j,2}$.  For the present consideration the explicit form of the $F_{j}$ is not relevant and for arbitrary $N$ is not particularly illuminating.

Now, let us briefly come back to the previous set of determining equations (\ref{quant deteq}) with $\ell=1$. This set of equations leads to the LCC (\ref{LCC00}) and, basically, they define all the functions
\begin{equation}\label{deffj2}
{f}_{j,2} \ = \ {f}_{j,2}(x,y,V,A_{N-m-n,\,m,\,n},N) \ ,
\end{equation}
in terms of $x,y$ and $V$. As a matter of fact, for $\ell=1$ one can solve (\ref{quant deteq}) straightforwardly. Therefore, substituting (\ref{deffj2}) into (\ref{FJg}) we can express the functions $F_j$ in terms of the potential $V$ (and its derivatives) only. Accordingly, the functions $F_j$ depend nonlinearly on the potential $V$. In this case, we can also obtain a nonlinear compatibility condition (see below).

For example, for $N=5$ we get from (\ref{quant deteq}) with $\ell=2$ (thus $j=0,1$) a system of equations of the form
\begin{equation}
\label{F4C}
\begin{aligned}
&   \partial_y{f}_{0,4} \ = \  F_0
\\ &
\partial_x{f}_{0,4} \ + \ \partial_y{f}_{1,4} \ = \  F_1
\\ &
\partial_x{f}_{1,4} \ = \  F_2  \ .
\end{aligned}
\end{equation}
We first solve (\ref{quant deteq}) with $\ell=1$ to obtain all the functions ${f}_{j,2}$. Then, in this case also $F_0$, $F_1$ and $F_2$ in (\ref{F4C}) are known. Hence, the compatibility condition of (\ref{F4C})
\[
\partial^2_{x}(\partial_y{f}_{0,4}) \ + \ \partial^2_{y}(\partial_x{f}_{1,4}) \ - \ \partial^2_{x,y}(\partial_x{ f}_{0,4}\,+\,\partial_y{ f}_{1,4}) \ \equiv \ 0 \ ,
\]
provides a nonlinear ODE for the potential $V$ alone (cf. eq. (28) in Ref 2).

Similarly in the case $N=6$, from (\ref{quant deteq}) with $\ell=1$ (thus $j=0,1,2$) we arrive to the following system of equations
\begin{equation}\label{}
\begin{aligned}
&   \partial_y{f}_{0,4} \ = \  F_0
\\ &
\partial_x{ f}_{0,4} \ + \ \partial_y{f}_{1,4} \ = \  F_1
\\ &
\partial_x{ f}_{1,4} \ + \ \partial_y{ f}_{2,4} \ = \  F_2
\\ &
\partial_x{ f}_{2,4} \ = \  F_3  \ .
\end{aligned}
\end{equation}
thus, their compatibility condition
\[
\partial_{y}\partial^2_{x}(\partial_x{f}_{04} + \partial_y{f}_{14} )\  - \ \partial_{x}\partial^2_{y}(\partial_x{ f}_{14} + \partial_y{ f}_{24}) \ - \ \partial^3_{x}(\partial_y{ f}_{04})  \ - \ \partial^3_{y}(\partial_x{ f}_{24})   \ \equiv \ 0 \ ,
\]
gives a nonlinear ODE for the potential $V$ as well.

Direct analysis of the determining equations (\ref{NLCCG}) with $\ell=2$, shows that for arbitrary odd $N>3$ (thus $j=0,1,2,\ldots,N-4$) we can obtain a NLCC for the potential $V$ as follows:

\begin{itemize}
  \item First, we solve the set of determining equations $M_{j,2\ell}=0$ (\ref{quant deteq}) with $\ell=1$. They define all the functions ${f}_{j,2}$ present in (\ref{YNQSd}).
  \item Then, from the next set of determining equations $M_{j,2\ell}=0$ with $\ell=2$ we calculate all the ($N-3$) functions ${f}_{j,4}$ except those with $j=\frac{N-5}{2}$ and $j=\frac{N-3}{2}$. That way, we arrive to the system
\begin{equation}\label{NLCCNODD}
\begin{aligned}
&   \partial_y{f}_{\frac{N-5}{2},4} \ = \  \tilde F_\frac{N-5}{2}
\\ &
\partial_x{ f}_{\frac{N-5}{2},4} \ + \ \partial_y{ f}_{\frac{N-3}{2},4} \ = \  \tilde F_\frac{N-3}{2}
\\ &
\partial_x{ f}_{\frac{N-3}{2},4} \ = \  \tilde F_\frac{N-1}{2}  \ ,
\end{aligned}
\end{equation}
here the $\tilde F$'s are real functions that depend on the potential $V$ (and its derivatives) only.
  \item Finally, the compatibility condition of (\ref{NLCCNODD})
\begin{equation}\label{Noddg}
\begin{aligned}
\partial^2_{x}(\partial_y{f}_{\frac{N-5}{2},4}) \ + \ \partial^2_{y}(\partial_x{ f}_{\frac{N-3}{2},4}) \ - \ \partial^2_{x,y}(\partial_x{ f}_{\frac{N-5}{2},4}\,+\,\partial_y{ f}_{\frac{N-3}{2},4}) \ \equiv \ 0 \ ,
\end{aligned}
\end{equation}
provides the aforementioned nonlinear ODE for the potential $V$.
\end{itemize}

As for $N>4$ odd, the procedure is similar

\begin{itemize}
  \item We solve the set of determining equations $M_{j,2\ell}=0$ with $\ell=1$. They define all the functions ${f}_{j,2}$.
  \item Then, from the next set of determining equations $M_{j,2\ell}=0$ with $\ell=2$ we calculate all the ($N-3$) functions ${f}_{j,4}$ except those with $j=\frac{N-6}{2},\frac{N-4}{2}$ and $j=\frac{N-2}{2}$. That way, we arrive to the system
\begin{equation}\label{NLCCNeven}
\begin{aligned}
&   \partial_y{ f}_{\frac{N-6}{2},4} \ = \   F_\frac{N-6}{2}
\\ &
\partial_x{ f}_{\frac{N-6}{2},4} \ + \ \partial_y{ f}_{\frac{N-4}{2},4} \ = \   F_\frac{N-4}{2}
\\ &
\partial_x{ f}_{\frac{N-4}{2},4} \ + \ \partial_y{ f}_{\frac{N-2}{2},4} \ = \   F_\frac{N-2}{2}
\\ &
\partial_x{ f}_{\frac{N-2}{2},4} \ = \  F_\frac{N}{2} \ ,
\end{aligned}
\end{equation}
here again the $F$'s are real functions that depend on the potential $V$ (and its derivatives) only.
  \item Hence, the compatibility condition of (\ref{NLCCNeven})
\begin{equation}\label{Neveng}
\begin{aligned}
& \partial_{y}\partial^2_{x}(\partial_x{ f}_{\frac{N-6}{2},4}\,+\,\partial_y{ f}_{\frac{N-4}{2},4})  \ - \
 \partial_{x}\partial^2_{y}(\partial_x{ f}_{\frac{N-4}{2},4}\,+\,\partial_y{ f}_{\frac{N-2}{2},4}) \ - \
\\ &
\partial^3_{x}(\partial_y{f}_{\frac{N-6}{2},4})  \ + \ \partial^3_{y}(\partial_x{f}_{\frac{N-2}{2},4})  \ \equiv \ 0 \ ,
\end{aligned}
\end{equation}
leads to a nonlinear ODE for the potential $V$.
\end{itemize}

\begin{figure}[htp]
  \centering
  \includegraphics[width=16.0cm]{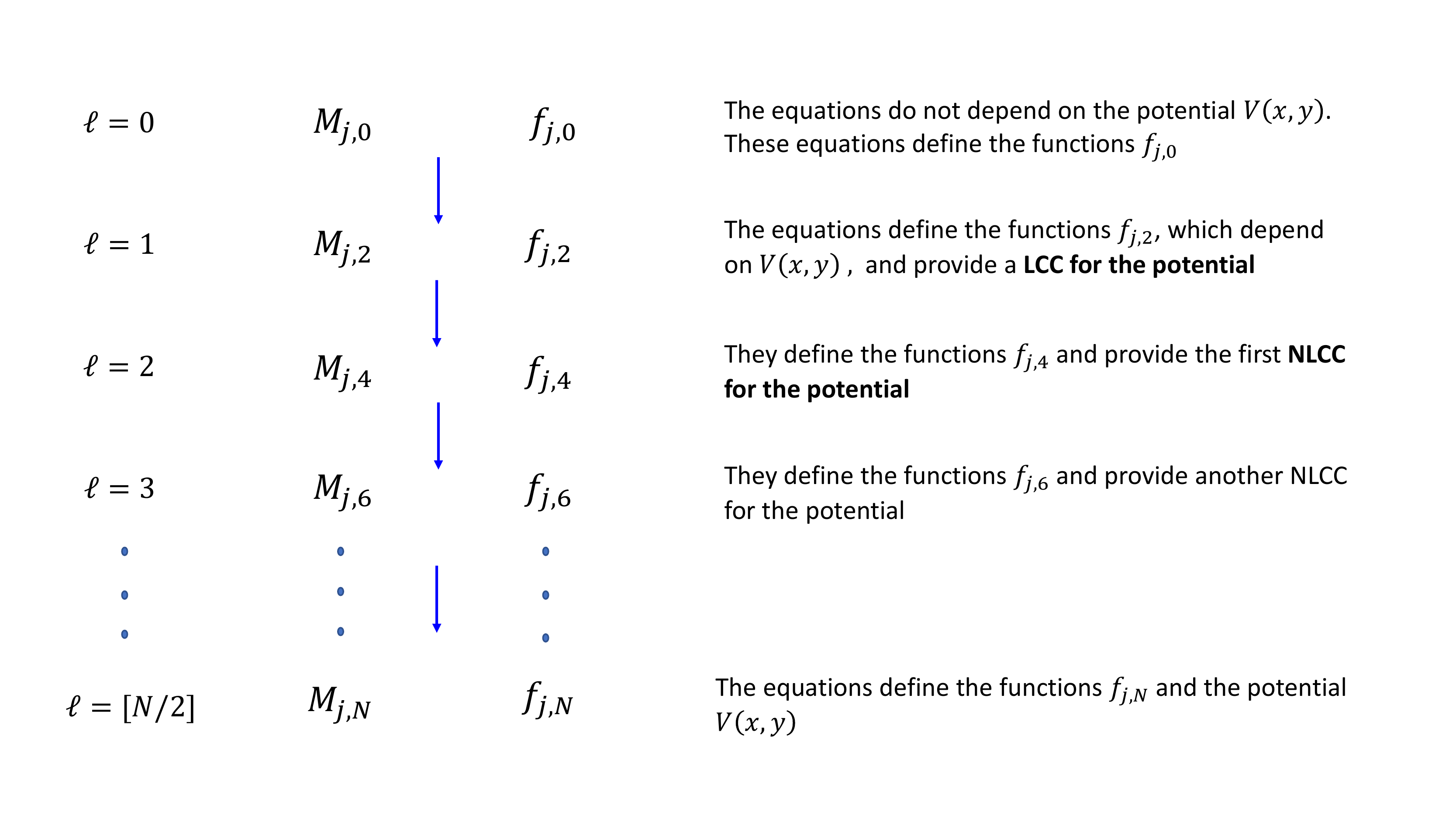}
  \caption{$N$th-order superintegrable systems. The condition $[H, \, Y_N]=0$ leads to a "well" of determining equations $M_{j,2\ell}=0$ (\ref{quant deteq}).}\label{Tab1}
\end{figure}

From the equations $M_{j,2\ell}=0$ (\ref{quant deteq}), it is clear that more NLCC occur with $\ell=3,4,\ldots,\big[\frac{N}{2} \big]$. These equations will restrict the general solution, found from the set $M_{j,4}=0$ ($\ell=2$), of the potential $V$ only. The general picture of the aforementioned procedure is summarized in Fig. \ref{Tab1}.

\clearpage

The entire procedure can be described as follows. The determining
equations (\ref{quant deteq}) which were originally
nonlinear PDEs are reduced to a coupled system of ODES by the separation
of variables. At the first level of the "well" the equations do not depend
on the potential and are hence linear ODEs that can be solved to obtain
the coefficients $f_{j,0}$. At the second level the equations for $f_{j,2}$ do depend on the potential but their compatibility condition is linear and separates into 2 linear ODEs for $V_1(x)$ and two for $V_2(y)$.
At all further levels the ODEs and their compatibility conditions are
nonlinear ODEs. The main object of interest are the compatibility
conditions since they determine the potential.

\clearpage

\section{Doubly exotic, singly exotic and standard potentials}
\label{families}

Based on the LCC (\ref{LCC00}), we define three major classes of superintegrable potentials:

\vspace{0.2cm}

\textbf{Doubly exotic potentials:} they satisfy the LCC (\ref{LCC00}) trivially. All four linear ODEs $\tau_{1}(x,V_1)=0$, $\tau_{2}(y,V_2)=0$, $\upsilon_{1}(x,V_1)=0$, and $\upsilon_{2}(y,V_2)=0$ (see (\ref{QX})-(\ref{QY})) vanish identically for any $V_1(x)$ and $V_2(y)$. Thus, the LCC (\ref{LCC00}) does not imply any linear ODE for $V_1(x)$ nor for $V_2(y)$. It follows that all the coefficients $A_{N-m-n,\,m,\,n}$ that figure in the LCC (\ref{LCC00}) must vanish but at least one of the other ones must survive so that the integral $Y_N$ (\ref{YNQSd}) is of order $N$.

\vspace{0.15cm}

\textbf{Singly exotic potentials:} a singly exotic potential $V_1(x)$ in $x$, occurs when both linear ODEs $\tau_{1}=0$ and $\upsilon_{1}=0$ in (\ref{QX}) are fulfilled trivially ($V_1(x)$ does not satisfy any linear ODE) while $V_2(y)$ obeys, non trivially, the LCC (\ref{LCC00}). Similarly, one can define singly exotic potentials $V_2(y)$ in $y$.
\vspace{0.15cm}

\textbf{Standard potentials:} they occur when neither $V_1(x)$ nor $V_2(y)$ are exotic, their most general forms are given by the solutions of the linear ODEs (\ref{QX}) and (\ref{QY}), respectively.

In this work we will focus on doubly exotic potentials.

\subsection{Integral $Y_N$ for doubly exotic potentials}

For doubly exotic (DE) potentials, i.e. those which satisfy the LCC (\ref{LCC00}) trivially, the Hermitian $N$th-order operator (\ref{YNQSd}) can be written in the form
\begin{equation}\label{YNDEQ}
Y_{N,DE} \ = \ W_N  \ + \ \frac{1}{2} \sum_{\ell=1}^{[\frac{N}{2}]}\sum_{j=0}^{N-2\ell}\,\{\,{\tilde f}_{j,2\ell}\,,\, p_1^j\ p_2^{N-j-2\ell}\,\} \qquad ; \qquad  \quad (N>2)  \ ,
\end{equation}
where
\begin{equation}\label{PNQ}
W_N \ = \ A_{0,N,0}\,p_1^N \ + \ A_{0,0,N}\,p_2^N  \ + \   \frac{1}{2} A_{N-4,2,2}\ \{\,L_z^{N-4}\,,\ p_1^2\,p_2^2\,\}
\end{equation}
\[
\, + \,  \frac{1}{2}\sum_{4 < m+n < N\,;\,\mid m-n\mid < N-4}^{}\ A_{N-m-n,m,n}\ \{\,L_z^{N-m-n}\,, \ p_1^m\,p_2^n\,\}  \, + \,  \sum_{0 \leq m+n = N\,;\,\mid m-n\mid \leq N-4}^{}\ A_{0,m,n}\ p_1^m\,p_2^n \ ,
\]
and $\{\,,\,\}$ denotes an anticommutator. The $N$th-order terms (\ref{YNDEQ}) contains $4(N-1)$ less parameters $A_{N-m-n,m,n}$ than the generic expression (\ref{YNQSd}). Notice that the angular momentum $L_z$ appears in the integral $Y_{N,DE}$ starting from $N=5$. For $N\geq5$, it can contain the powers $L_z, L_z^2, L_z^3,\ldots ,L_z^{N-4}$ only. For large $N$, the number of terms in $W_N$ grows as $N^2$. Below, for the lowest cases $N=3,4,\ldots,10$ we present the most general leading term $W_N$ of the integral $Y_{N,DE}$ explicitly

\begin{equation}\label{YN3de}
W_3  \ =  \ A_{030}\,p_1^3  \ + \  A_{003}\,p_2^3  \ .
\end{equation}

\begin{equation}\label{YN4de}
W_4  \ =  \ A_{040}\,p_1^4  \ + \  A_{004}\,p_2^4 \ + \  A_{022}\,p_1^2 \,p_2^2 \ .
\end{equation}

\begin{equation}\label{YN5de}
W_5  \ =  \ A_{050}\,p_1^5  \ + \  A_{005}\,p_2^5\ + \  A_{032}\,p_1^3\,p_2^2  \ + \  A_{023}\,p_1^2\,p_2^3 \ + \  \frac{1}{2}A_{122}\{L_z,\,p_1^2 \,p_2^2\,\} \ .
\end{equation}

\begin{equation}\label{YN6de}
\begin{aligned}
W_6  \  = & \ A_{060}\,p_1^6  \ + \  A_{006}\,p_2^6 \ + \  A_{033}\,p_1^3 \,p_2^3\ + \  A_{024}\,p_1^2 \,p_2^4 \ + \  A_{042}\,p_1^4 \,p_2^2
\\ & \ + \  \frac{1}{2}A_{222}\,\{L_z^{2},\,p_1^2 \,p_2^2\} \ + \  \frac{1}{2}A_{123}\,\{ L_z,\,p_1^2 \,p_2^3\}\ + \  \frac{1}{2}A_{132}\,\{L_z,\,p_1^3 \,p_2^2\} \ .
\end{aligned}
\end{equation}

\begin{equation}\label{YN7de}
\begin{aligned}
W_7  \ =  & \ A_{070}\,p_1^7  \ + \  A_{007}\,p_2^7\ + \  \frac{1}{2}A_{322}\,\{L_z^3,\,p_1^2 \,p_2^2\,\} \ + \  \frac{1}{2}A_{232}\,\{L_z^2,\,p_1^3 \,p_2^2\,\} \ + \ \frac{1}{2}A_{223}\,\{L_z^2,\,p_1^2 \,p_2^3\,\} \\ &
\ + \ \frac{1}{2}A_{142}\,\{L_z,\,p_1^4 \,p_2^2\,\} \ + \ \frac{1}{2}A_{124}\,\{L_z,\,p_1^2 \,p_2^4\,\}
\ + \ \frac{1}{2}A_{133}\,\{L_z,\,p_1^3 \,p_2^3\,\} \\ &
 \ + \  A_{052}\,p_1^5 \,p_2^2 \ + \  A_{025}\,p_1^2 \,p_2^5
 \ + \  A_{043}\,p_1^4 \,p_2^3 \ + \  A_{034}\,p_1^3 \,p_2^4  \ .
\end{aligned}
\end{equation}

\begin{equation}\label{YN8de}
\begin{aligned}
W_8  \ =  & \ A_{080}\,p_1^8  \ + \  A_{008}\,p_2^8\ + \  \frac{1}{2}A_{422}\,\{L_z^4,\,p_1^2 \,p_2^2\,\} \ + \  \frac{1}{2}A_{323}\,\{L_z^3,\,p_1^2 \,p_2^3\,\} \ + \ \frac{1}{2}A_{332}\,\{L_z^3,\,p_1^3 \,p_2^2\,\} \\ &
\ + \ \frac{1}{2}A_{242}\,\{L_z^2,\,p_1^4 \,p_2^2\,\} \ + \ \frac{1}{2}A_{224}\,\{L_z^2,\,p_1^2 \,p_2^4\,\}
\ + \ \frac{1}{2}A_{233}\,\{L_z^2,\,p_1^3 \,p_2^3\,\} \\ &
\ + \ \frac{1}{2}A_{152}\,\{L_z,\,p_1^5 \,p_2^2\,\} \ + \ \frac{1}{2}A_{125}\,\{L_z,\,p_1^2 \,p_2^5\,\}
\ + \ \frac{1}{2}A_{143}\,\{L_z,\,p_1^4 \,p_2^3\,\} \ + \ \frac{1}{2}A_{134}\,\{L_z,\,p_1^3 \,p_2^4\,\} \\ &
 \ + \  A_{062}\,p_1^6 \,p_2^2 \ + \  A_{026}\,p_1^2 \,p_2^6
 \ + \  A_{035}\,p_1^3 \,p_2^5 \ + \  A_{053}\,p_1^5 \,p_2^3
 \\ &
  \ + \  A_{044}\,p_1^4 \,p_2^4 \ .
\end{aligned}
\end{equation}

\begin{equation}\label{YN9de}
\begin{aligned}
W_9  \ =  & \ A_{090}\,p_1^9  \ + \  A_{009}\,p_2^9\ + \  \frac{1}{2}A_{522}\,\{L_z^5,\,p_1^2 \,p_2^2\,\} \ + \  \frac{1}{2}A_{423}\,\{L_z^4,\,p_1^2 \,p_2^3\,\} \ + \ \frac{1}{2}A_{432}\,\{L_z^4,\,p_1^3 \,p_2^2\,\} \\ &
\ + \ \frac{1}{2}A_{342}\,\{L_z^3,\,p_1^4 \,p_2^2\,\} \ + \ \frac{1}{2}A_{324}\,\{L_z^3,\,p_1^2 \,p_2^4\,\}
\ + \ \frac{1}{2}A_{333}\,\{L_z^3,\,p_1^3 \,p_2^3\,\} \\ &
\ + \ \frac{1}{2}A_{252}\,\{L_z^2,\,p_1^5 \,p_2^2\,\} \ + \ \frac{1}{2}A_{225}\,\{L_z^2,\,p_1^2 \,p_2^5\,\}
\ + \ \frac{1}{2}A_{243}\,\{L_z^2,\,p_1^4 \,p_2^3\,\} \ + \ \frac{1}{2}A_{234}\,\{L_z^2,\,p_1^3 \,p_2^4\,\} \\ &
\ + \ \frac{1}{2}A_{162}\,\{L_z,\,p_1^6 \,p_2^2\,\} \ + \ \frac{1}{2}A_{126}\,\{L_z,\,p_1^2 \,p_2^6\,\}
\ + \ \frac{1}{2}A_{153}\,\{L_z,\,p_1^5 \,p_2^3\,\} \ + \ \frac{1}{2}A_{135}\,\{L_z,\,p_1^3 \,p_2^5\,\} \\ &
\ + \ \frac{1}{2}A_{144}\,\{L_z,\,p_1^4 \,p_2^4\,\}  \ + \  A_{072}\,p_1^7 \,p_2^2 \ + \  A_{027}\,p_1^2 \,p_2^7
 \ + \  A_{045}\,p_1^4 \,p_2^5 \ + \  A_{054}\,p_1^5 \,p_2^4
 \\ &
  \ + \  A_{036}\,p_1^3 \,p_2^6 \ + \  A_{063}\,p_1^6 \,p_2^3    \ .
\end{aligned}
\end{equation}

\begin{equation}\label{YN10de}
\begin{aligned}
W_{10}  \ =  & \ A_{0,10,0}\,p_1^{10}  \ + \  A_{0,0,10}\,p_2^{10}\ + \  \frac{1}{2}A_{622}\,\{L_z^6,\,p_1^2 \,p_2^2\,\} \ + \  \frac{1}{2}A_{523}\,\{L_z^5,\,p_1^2 \,p_2^3\,\} \ + \ \frac{1}{2}A_{532}\,\{L_z^5,\,p_1^3 \,p_2^2\,\} \\ &
\ + \ \frac{1}{2}A_{442}\,\{L_z^4,\,p_1^4 \,p_2^2\,\} \ + \ \frac{1}{2}A_{424}\,\{L_z^4,\,p_1^2 \,p_2^4\,\}
\ + \ \frac{1}{2}A_{433}\,\{L_z^4,\,p_1^3 \,p_2^3\,\} \\ &
\ + \ \frac{1}{2}A_{352}\,\{L_z^3,\,p_1^5 \,p_2^2\,\} \ + \ \frac{1}{2}A_{325}\,\{L_z^3,\,p_1^2 \,p_2^5\,\}
\ + \ \frac{1}{2}A_{343}\,\{L_z^3,\,p_1^4 \,p_2^3\,\} \ + \ \frac{1}{2}A_{334}\,\{L_z^3,\,p_1^3 \,p_2^4\,\} \\ &
\ + \ \frac{1}{2}A_{262}\,\{L_z^2,\,p_1^6 \,p_2^2\,\} \ + \ \frac{1}{2}A_{226}\,\{L^2_z,\,p_1^2 \,p_2^6\,\}
\ + \ \frac{1}{2}A_{163}\,\{L_z,\,p_1^6 \,p_2^3\,\} \ + \ \frac{1}{2}A_{136}\,\{L_z,\,p_1^3 \,p_2^6\,\}
\\ &
\ + \ \frac{1}{2}A_{172}\,\{L_z,\,p_1^7 \,p_2^2\,\} \ + \ \frac{1}{2}A_{127}\,\{L_z,\,p_1^2 \,p_2^7\,\}
\ + \ \frac{1}{2}A_{154}\,\{L_z,\,p_1^5 \,p_2^4\,\} \ + \ \frac{1}{2}A_{145}\,\{L_z,\,p_1^4 \,p_2^5\,\} \\ &
 \ + \  A_{028}\,p_1^2 \,p_2^8 \ + \  A_{082}\,p_1^8 \,p_2^2  \ + \  A_{037}\,p_1^3 \,p_2^7 \ + \  A_{073}\,p_1^7 \,p_2^3
 \\ &
  \ + \  A_{046}\,p_1^4 \,p_2^6 \ + \  A_{064}\,p_1^6 \,p_2^4  \ + \  A_{055}\,p_1^5 \,p_2^5     \ .
\end{aligned}
\end{equation}

However, this number can be effectively reduced from the very beginning due to the existence of trivial higher order integrals (see below.)

\bigskip

\section{Trivial integrals}
\label{trivint}

Two types of trivial higher order integrals exist.

\subsection{Integrals of even order $N$ that are consequences of second order integrability}

Integrals of even order $N=2\,K$ ($K \in \mathbb{N}$) for arbitrary separable systems (\ref{Hcart}), (\ref{X}) exist. They are of the form

\begin{equation}\label{YTEV}
Y_N^{Tr} \ = \ \sum_{0\leq i+j \leq K}^{}\,\alpha_{ij}\,X^i\,H^j \ + \ l.o.t. \ ,
\end{equation}

where $\alpha_{ij} \in \mathbb{R}$ are arbitrary constants. The highest order terms $W_N^{Tr}$ (\ref{PNQ}) of the integral $Y_N^{Tr}$ can be rewritten as

\begin{equation}\label{PNTr}
  W_N^{Tr} \ = \ \sum_{j=0}^{K} \,\alpha_j\,p_1^{2(K-j)}\,p_2^{2j} \ .
\end{equation}

When searching for superintegrable systems $(H,X,Y_N)$ we can simplify from the very beginning by linear combinations of $Y_N$ with $Y_N^{Tr}$ and use the constants $\alpha_j$ in (\ref{PNTr}) to annihilate all the terms of the form $L_z^0\,p_1^{2(k-j)}\,p_2^{2j}$, i.e. we put
\begin{equation}\label{}
  A_{0,\,2(k-j),\,2j} \ = \ 0  \quad , \qquad j=0,1,2\ldots,k \ ,
\end{equation}
in $W_N$ (\ref{PNQ}).

For even $N\geq 4$ this greatly restricts the number of doubly exotic systems. E.g. for $N=4$ (see (\ref{YN4de})) no doubly exotic systems exist since $A_{040}$, $A_{022}$ and $A_{004}$ all correspond to trivial integrals. For $N=6$ the only nontrivial integral not involving $L_z$ would correspond to $A_{033}$. However, even this case can be excluded since the square of the corresponding integral can be expressed in terms of trivial integrals (a syzygy)

\begin{equation}\label{}
  {(p_1^3\,p_2^3)}^2 \ = \ (p_1^4\,p_2^2)(p_1^2\,p_2^4) \ .
\end{equation}

Similarly, for $N=8$ doubly exotic potentials (\ref{YN8de}) the terms $A_{080}$, $A_{026}$, $A_{044}$, $A_{062}$ and $A_{008}$ all correspond to trivial integrals. The remaining two integrals not involving $L_z$ correspond to $A_{035}$ and $A_{053}$, and again they can be expressed nonlinearly in terms of products of trivial integrals

\begin{equation}\label{}
   {(p_1^5\,p_2^3)}^2 \ = \ p_1^8\,(p_1^2\,p_2^6) \qquad , \qquad  {(p_1^3\,p_2^5)}^2 \ = \ p_2^8\,(p_2^2\,p_1^6) \ .
\end{equation}

\textbf{Remark.} Since there exists a syzygy with trivial integrals $H$ and $X$, in the case of even order $N=2\,K$ ($K \in \mathbb{N}$) all the terms in $Y_{2K}$ of the form $p_1^{2K-a}\,p_2^a$ and $p_2^{2K-a}\,p_1^a$ can be discarded from the very beginning. In particular, this implies that $W_4$ is identically zero while $W_6$, $W_8$ and $W_{10}$ reduce to

\begin{equation}\label{YN6der}
\begin{aligned}
W_6  \  = &  \frac{1}{2}A_{222}\,\{L_z^{2},\,p_1^2 \,p_2^2\} \ + \  \frac{1}{2}A_{123}\,\{ L_z,\,p_1^2 \,p_2^3\}\ + \  \frac{1}{2}A_{132}\,\{L_z,\,p_1^3 \,p_2^2\} \ .
\end{aligned}
\end{equation}

\begin{equation}\label{YN8der}
\begin{aligned}
W_8  \ =  & \ \frac{1}{2}A_{422}\,\{L_z^4,\,p_1^2 \,p_2^2\,\} \ + \  \frac{1}{2}A_{323}\,\{L_z^3,\,p_1^2 \,p_2^3\,\} \ + \ \frac{1}{2}A_{332}\,\{L_z^3,\,p_1^3 \,p_2^2\,\} \\ &
\ + \ \frac{1}{2}A_{242}\,\{L_z^2,\,p_1^4 \,p_2^2\,\} \ + \ \frac{1}{2}A_{224}\,\{L_z^2,\,p_1^2 \,p_2^4\,\}
\ + \ \frac{1}{2}A_{233}\,\{L_z^2,\,p_1^3 \,p_2^3\,\} \\ &
\ + \ \frac{1}{2}A_{152}\,\{L_z,\,p_1^5 \,p_2^2\,\} \ + \ \frac{1}{2}A_{125}\,\{L_z,\,p_1^2 \,p_2^5\,\}
\ + \ \frac{1}{2}A_{143}\,\{L_z,\,p_1^4 \,p_2^3\,\} \ + \ \frac{1}{2}A_{134}\,\{L_z,\,p_1^3 \,p_2^4\,\} \ .
\end{aligned}
\end{equation}

\begin{equation}\label{YN10der}
\begin{aligned}
W_{10}  \ =  & \  \frac{1}{2}A_{622}\,\{L_z^6,\,p_1^2 \,p_2^2\,\} \ + \  \frac{1}{2}A_{523}\,\{L_z^5,\,p_1^2 \,p_2^3\,\} \ + \ \frac{1}{2}A_{532}\,\{L_z^5,\,p_1^3 \,p_2^2\,\} \\ &
\ + \ \frac{1}{2}A_{442}\,\{L_z^4,\,p_1^4 \,p_2^2\,\} \ + \ \frac{1}{2}A_{424}\,\{L_z^4,\,p_1^2 \,p_2^4\,\}
\ + \ \frac{1}{2}A_{433}\,\{L_z^4,\,p_1^3 \,p_2^3\,\} \\ &
\ + \ \frac{1}{2}A_{352}\,\{L_z^3,\,p_1^5 \,p_2^2\,\} \ + \ \frac{1}{2}A_{325}\,\{L_z^3,\,p_1^2 \,p_2^5\,\}
\ + \ \frac{1}{2}A_{343}\,\{L_z^3,\,p_1^4 \,p_2^3\,\} \ + \ \frac{1}{2}A_{334}\,\{L_z^3,\,p_1^3 \,p_2^4\,\} \\ &
\ + \ \frac{1}{2}A_{262}\,\{L_z^2,\,p_1^6 \,p_2^2\,\} \ + \ \frac{1}{2}A_{226}\,\{L^2_z,\,p_1^2 \,p_2^6\,\}
\ + \ \frac{1}{2}A_{163}\,\{L_z,\,p_1^6 \,p_2^3\,\} \ + \ \frac{1}{2}A_{136}\,\{L_z,\,p_1^3 \,p_2^6\,\}
\\ &
\ + \ \frac{1}{2}A_{172}\,\{L_z,\,p_1^7 \,p_2^2\,\} \ + \ \frac{1}{2}A_{127}\,\{L_z,\,p_1^2 \,p_2^7\,\}
\ + \ \frac{1}{2}A_{154}\,\{L_z,\,p_1^5 \,p_2^4\,\} \ + \ \frac{1}{2}A_{145}\,\{L_z,\,p_1^4 \,p_2^5\,\}     \ ,
\end{aligned}
\end{equation}

respectively.

\subsection{Integrals of order $N$ that are consequences of lower order superintegrability}

Let us consider the case of a Hamiltonian system of the type (\ref{Hcart}), (\ref{X}) that is already superintegrable for some potential $V(x,y)=V_1(x)+V_2(y)$. Then in addition to the integrals $H$ and $X$ we have a further nontrivial integral $Y_M$ of the form (\ref{YNQSd}). In a two-dimensional space $E_2$ we can have at most 3 algebraically independent integrals. Thus, there is no need to search for further integrals for this potential.

On the other hand, if we wish to find all superintegrable systems of order $N>M$ (in particular all doubly exotic ones) it is necessary to discard all those that are really lower order superintegrable and were already known. For those the $N$th-order integral will be reducible. In this case $Y_N$ will be a polynomial in $Y_M$, $H,X$ and possibly commutators of the type $[Y_M^p, X^q]$ where $p$ and $q$ are positive integers.
These commutators will all vanish unless the leading term in $Y_M$ involves the operator $L_z$ to some positive integer power.

\textbf{Remark.} For odd $N$, if $W_N$ is a polynomial in $W_M$,$H,X$ along with the trivial (lower order) integral $Y_W$, a more general $N$th-order superintegrable potential can occur. Thus, in general, such terms can not be discarded from the very beginning.

\section{Two infinite families of doubly exotic potentials}
\label{partcases}

The Hamiltonian $H$ and the integral $X$ are invariant and anti-invariant, respectively, under the action of the permutation group $\mathcal{S}_2$, i.e. $x\Leftrightarrow y$ (thus, $p_1 \Leftrightarrow p_2$). Therefore, it is natural to consider the $N$th-order integral $Y_N$ being either $\mathcal{S}_2$ invariant or anti-invariant. In this study, we will restrict ourselves to the following two cases:

\begin{itemize}
  \item Integrals of the form $Y^{(I)}_{2K+1} \ = \  \cos \gamma\,\,p_1^{2K+1} \ + \ \sin \gamma\,\,p_2^{2K+1} \ + \ l.o.t \ $, where $0< K \in \mathbb{N}$ and $0 < \gamma \neq \frac{\pi}{2} < \pi$. In this case, the integral $Y^{(I)}_{2K+1}$ is ${\mathcal{S}}_2$-invariant.
  \item Integrals of the form $Y^{(II)}_N \ = \  \{L_z^{N-4},\,p_1^{2} \,p_2^{2}\,\} \ + \ l.o.t \ $, where $N>5$. Hence, for $N$ even $Y_N$ is invariant while for $N$ odd it is anti-invariant.
\end{itemize}

By doing so, we are aware that a number of doubly exotic potentials will be overlooked. Here we announce main findings only, with a detailed complete classification in papers under preparation. Moreover, within the present approach those overlooked systems can be easily obtained as well.

Hereafter, we will be mainly interested in the two above mentioned families of doubly exotic $N$th-order superintegrable systems with integrals for which the highest order terms have the form

\begin{equation}\label{PN1T}
 W_{N}(I) \ = \ \cos \gamma \,p_1^N \ + \ \sin \gamma \,p_2^N    \qquad ; \qquad N \, = \,2K+1  \ , \qquad K > 0 \ ,
\end{equation}

$0 < \gamma \neq \frac{\pi}{2}< \pi $, or

\begin{equation}\label{PN2T}
 W_{N}(II) \ = \ \{\,L_z^{(N-4)},\,p_1^2\,p_2^2    \, \}   \qquad ;  \qquad N > 4 \ .
\end{equation}

It is worth mentioning that in Refs.\cite{MSW2,GungorKNN:2014} an elegant systematic use of algebraic systems in one
dimension is exploited to generate 2D superintegrable systems. In particular, the two cases $W_{N}(I)$ and $W_{N}(II)$ correspond to the types $(b,b)$ and $(d,d)$, respectively (see Table I in Ref.\cite{MSW2}). As for the algebraic approach see the papers\cite{Boyer,Doebner,Hietarinta1987,Iliev,IanMcubic,IanMladder,Veselov}.
However, it is not clear if all 2D superintegrable potentials can be obtained that way. Also, the analogue of the general linear equation (\ref{LCC00}) which in the present formalism plays a fundamental role to define the class of superintegrable potentials, especially the doubly exotic ones, is absent. In this sense, the two methods complement each other.

Now, as a first step we need to show that the integrals (\ref{PN1T}) and (\ref{PN2T}) are not trivial ones (irreducibility). We shall consider the two cases separately.

\subsubsection{$W_N(I)$}

A polynomial of the form (\ref{PN1T}) can never have the form (\ref{YTEV}) (for $N$ odd). The question is whether $W_N(I)$ can be obtained from a lower order $W_M$, $N>M$. The potential functions $V_1$ and $V_2$ must satisfy the same nonlinear equations for the values $M$ and $N$.

Let us first assume that $W_M=W_M(I)$, i.e. $W_M$ and $W_N(I)$ belong to the same family. For $W_N(I)\neq 0$ to be a trivial extension of $W_M(I)$ we must have ($N=2\,K+M$)

\begin{equation}\label{}
\begin{aligned}
  W_N \ & = \ W_M\,W_{2K}^{Tr} \ = \ (\cos \gamma \,p_1^M \ + \ \sin \gamma \,p_2^M )\sum_{j=0}^{K} \,\alpha_j\,p_1^{2(K-j)}\,p_2^{2j}
  \\ &
   \ = \ \cos \gamma \sum_{j=0}^{K} \,\alpha_j\,p_1^{2(K-j)+M}\,p_2^{2j} \ + \ \sin \gamma \sum_{j=0}^{K} \,\alpha_j\,p_1^{2(K-j)}\,p_2^{2j+M}
\end{aligned}
\end{equation}

For $W_N$ to be in the family (\ref{PN1T}) we must annihilate all terms other than monomials in $p_1$ or in $p_2$. For $\cos \gamma \neq 0$, this implies $\alpha_j = \alpha_0\,\delta_{j,0}$  or $\alpha_j = \alpha_{M}\,\delta_{M,2(j-K)}$. For $\sin \gamma \neq 0$, this implies $\alpha_j = \alpha_{K}\,\delta_{j,K}$. Hence, $W_{2K}^{Tr}=0$ (then $W_N=0$) which is a contradiction. \quad $\Box$

Other integrals of order $M$ involving only the momenta $p_1$ and $p_2$, not belonging to the family (\ref{PN1T}) also come in pairs

\begin{equation}\label{}
  W_M \ = \ \sum_{j=1}^{M-1}\,\big[\,A_{0,M-j,j}\,p_1^{M-j}\,p_2^j \ + \ A_{0,j,M-j}\,p_1^{j}\,p_2^{M-j}  \,\big] \ .
\end{equation}

Multiplying the above $W_M$ by a trivial integral (\ref{PNTr}) we obtain again an integral that can not be reduced to the form (\ref{PN1T}) for any choice of the arbitrary parameters $\alpha_j$. Hence, integrals of the form (\ref{PN1T}) are irreducible.

\subsubsection{$W_N(II)$}

Finally an integral of the form (\ref{PN2T}) with leading term containing $L_z^2$ can be generated by lower order integral of the form $L_z^{2K}\,p_1^A\,p_2^B$ by commutations with a trivial integral $Y_{2m}^{Tr} ={(p_1^2 - p_2^2)}^m$. This can be reduced to the case

\begin{equation}\label{}
   [\,  L_z^{2K}\,p_1^A\,p_2^B, \,  {p_1^2-p_2^2} \,] \ = \ 4\,L_z^{2K-1}p_1^{A+1}p_2^{B+1} \ + \ l.o.t. \ .
\end{equation}

so, products of $p_1$ and $p_2$ are inevitable. Thus, e.g. we can obtain a trivial integral of order 7 from a nontrivial one of order 6. For $Y_6$, the integral with $A_{222}\,\{\, L_z^2,\,p_1^2\,p_2^2  \, \}$ will yield a trivial integral of the form $Y_7 \, =\, A_{133}\,\{\, L_z,\,p_1^3\,p_2^3  \, \} + l.o.t.$, but never $\{\, L_z^3,\,p_1^2\,p_2^2  \, \}$.

More generally, the commutation will decrease the power of $L_z$ by one, never increase it. Therefore, integrals of the form (\ref{PN2T}) are irreducible as well.

\bigskip

\section{Low order examples $(3\leq N \leq 10)$}
\label{DoublyExoticlow}

The leading terms of the integral $Y_N$ were given in Section VI in eqs. (\ref{YN3de})-(\ref{YN10de}). Here we shall treat the cases $N=3,4,\dots,10$ in more detail. In all cases we will respect the $x \Leftrightarrow y$ permutation symmetry. With respect to
this symmetry the expressions for $W_N$ contain singlets (like $L_z^a\, p_1^b\,
p_2^b)$ and doublets (like $L_z^a\,  p_1^b\,  p_2^c$,  $L_z^a\, p_1^c\, p_2^b$ with $b\neq c$).
When solving the determining equations for a given value of $N$ we shall
simplify $W_N$ by eliminating all trivial integrals and restricting to just
one singlet or to a single doublet. This will not necessarily give us a
complete classification of all doubly exotic potentials. It will however
enable us to find many examples and to confirm the existence of the two families
$W_N(I)$ (\ref{PN1T}) and $W_N(II)$ (\ref{PN2T}) of superintegrable systems. For all cases considered we have obtained
nonlinear ODEs for the potential functions and we have shown that they
pass the Painlev\'e test. For $N=3$ all doubly exotic systems are known\cite{gW}
and the ODEs have been integrated in terms of the original Painlev\'e
transcendents (so they do actually have th Painlev\'e property\cite{Bureau1,Bureau2,Cosgrove,Cosgrove:SD}). For $N=4$
there are no doubly exotic potentials since all fourth order integrals
trivial (products of second order ones. For $N= 5$ all doubly exotic
potentials are also known\cite{AW} and all of the ODEs for the potential
functions pass the Painlev\'e test. Most of them have been integrated in
terms of the known (second order) Painlev\'e transcendents. Others probably
define new (higher order) transcendents. Unfortunately it is difficult to
prove that the corresponding Laurent series have a nonzero radius of
convergence. For $N>5$ all results presented below are new.

\subsection{Case $ N=3$}

\bigskip

For $N=3$, there exists only one doubly exotic superintegrable potential. The corresponding integral is a doublet
with respect to the $x,y$ permutation and is given by

\vspace{0.2cm}

$ \bullet \ Y_3 \ = \ \cos \gamma\,p_1^3  \ + \ \sin \gamma\,p_2^3 \ + \ \ldots$

\vspace{0.4cm}

\begin{equation}\label{VN3}
V(x,\,y) \ = \  \hbar^2 \,\big[ \omega_1^2\,P_1(\omega_1\,x) \ + \ \omega_2^2\,P_1(\omega_2\,y)\, \big] \ ,
\end{equation}
($0 < \gamma \neq \frac{\pi}{2}< \pi $, $\omega_1^5 = \cos \gamma$, $\omega_2^5 =-\sin \gamma$ ) where $P_1=P_1(u)$ satisfies the first Painlev\'e equation
\[
P_1'' \ = \ 6\,P_1^2 \ + \ u \ .
\]

This result was obtained in Ref.\cite{gW} (eq. $(Q.17)$) and Ref.\cite{GungorKNN:2014} (eq. (3.22)-(3.23)), independently. This is the first element of an infinite family ($W_N(I)$) of higher order superintegrable potentials separating in Cartesian coordinates. This family is characterized by an $N$th-order integral ${\cal Y}_N$ separating in Cartesian coordinates, that is, ${\cal Y}_N = {Y}_N(x,p_1) \, + \, {Y}_N(y,p_2)$.

\subsection{Case $ N=4$}

In this case, all the fourth order terms (\ref{YN4de}) in $Y_4$ can be written as a polynomial in the trivial integrals $H$ and $X$. The corresponding fourth order integral would correspond to one of the
original second order superintegrable potentials introduced in Refs.\cite{Fris:1965, Evans1991}, namely
$V=\omega\, (x^2 +y^2)+\frac{\beta}{x^2} + \frac{\gamma}{y^2}$. Therefore, no doubly exotic potentials with a non trivial fourth order integral exist.

\subsection{Case $ N=5$}

\bigskip

The decision to respect the $(x,y)$ permutation symmetry restricts the $N=5$ case to one singlet and two doublets treated below.

\vspace{0.2cm}

$ \bullet \ Y_5^{(a)} \ = \ \{\,L_z,\,p_1^2 \,p_2^2\} \ + \  \ldots$
\vspace{0.4cm}

The potential is given by
\begin{equation}\label{}
V(x,y)\ = \ \hbar^2\,[\,\mathcal{F}'(x;a,b) \ + \ \mathcal{F}'(y;\tilde a,\tilde b)\,] \ ,
\end{equation}
where the function $\mathcal{F}=\mathcal{F}(z;a,b)$ satisfies a nonlinear ODE of the form
\begin{equation}\label{N5A122}
\begin{aligned}
& {\mathcal{F}}^{(3)}  \ - \
 {\mathcal{F}}'\,\big[\,6\,{\mathcal{F}}' \,+\, a\,+\,\sigma\,z^2   \,\big]
  \ - \ \,2\,\sigma\,z\,{\mathcal{F}}\ + \ b_0  \ + \ b_1\,z\ + \ b_2\,z^2 \ + \   \Lambda\,z^4  \ = \ 0 \ ,
\end{aligned}
\end{equation}
$\mathcal{F}^{(\ell)} \equiv \frac{d^\ell}{dz^\ell}\,\mathcal{F}(z)$, here $a$, $b_k$ ($k=0,1,2$), $\sigma$ and $\Lambda$ are real constants. For $\sigma \neq 0$, up to a redefinition of the constants, the solutions of (\ref{N5A122}) are given by

\[
2\,V_1(x) \ = \  \hbar^2 \bigg(\, \sqrt{\alpha}\,P_4' \ - \ \alpha\,x\,P_4 \ - \ \alpha\,P_4^2  \ - \ \alpha\,\frac{x^2}{4}\,\bigg) \ ,
\]
\begin{equation}\label{}
2\,V_2(y) \ = \  \hbar^2 \bigg(\, \sqrt{\alpha}\,P_4' \ - \ \alpha\,y\,P_4 \ - \ \alpha\,P_4^2  \ - \ \alpha\,\frac{y^2}{4}\,\bigg) \ ,
\end{equation}
where $P_4=P_4(u;\,\alpha \neq 0,\,K_1,\,K_2)$ satisfies the fourth Painlev\'e equation
\begin{equation}
	P_4''\ = \ \frac{\left(P_4'\right){}^2}{2\, P_4}\ - \ \frac{3}{2}\alpha \,P_4^3\ - \ 2\,\alpha \,u\, P_4^2\ - \ \bigg(\frac{1}{2}\alpha \,u^2+K_1\bigg)\,P_4\ + \ \frac{K_2}{P_4}\ .
	\end{equation}
(see eq. $(139)$ in Ref.\cite{AW}). This potential starts the second infinite family $W_N(II)$ of higher order superintegrable potentials  separating in Cartesian coordinates.

\bigskip

\vspace{0.4cm}

$\bullet \ Y_5^{(b)} \ = \ \cos \gamma \, p_1^5 \ + \ \sin \gamma \,p_2^5 \ + \  \ldots$

\vspace{0.4cm}

This doublet is the second term in the family $I$ starting at $N=3$. For the function $V_1$ we obtain the fourth order ODE

\begin{align}
\label{V1050}
	& \cos \gamma\,\bigg[ \,a_0  \ + \ a_2\,V_1  \ + \ 6\,a_1\,V_1^2 \ + \ 40\,V_1^3 \ - \ a_1\,\hbar^2\,V_1''
\ - \ 10\,\hbar^2\,{(V_1')}^2  \nonumber \\ & \ - \ 20\,\hbar^2\,V_1\,V_1'' \ + \ \hbar^4\,V_1^{(4)} \,\bigg] \ = \ \lambda\,x   \ ,
\end{align}

functionally identical to the one obtained for the function $V_2$

\begin{align}
\label{}
	& \sin \gamma\,\bigg[\,b_0  \ + \ b_2\,V_2  \ + \ 6\,b_1\,V_2^2 \ + \ 40\,V_2^3 \ - \ b_1\,\hbar^2\,V_2''
\ - \ 10\,\hbar^2\,{(V_2')}^2  \nonumber \\ & \ - \ 20\,\hbar^2\,V_2\,V_2'' \ + \ \hbar^4\,V_2^{(4)}\bigg] \ = \ - \lambda\,y  \ ,
\end{align}

Here $a_k$, $b_k$ are arbitrary real parameters. Again, the above results were found in Ref.\cite{AW} (eq. $(146)$) and Ref.\cite{GungorKNN:2014} (eq. (3.30)), independently. The above two equations pass the Painlev\'e test for $\hbar \neq 0$ only. The resonances occur at $r=2,5,8$. For particular cases of the parameters, they appear in the list of fourth order Painlev\'e equations of polynomial class, classified by Cosgrove \cite{Cosgrove}: it is precisely the so called equation F-V (see equation (2.67) in\cite{Cosgrove} with $\alpha=\beta=0$). It is conjectured that these equations define a new transcendent in the sense that their general solution cannot be expressed in terms of the six Painlev\'e transcendents.

A complete analysis of the doubly exotic potentials for $N=5$ was
presented earlier\cite{AW}. In addition to the cases presented above several
other ones were obtained corresponding to mixed non symmetric integrals
like $Y_5 =A_{023}\, p_1^2 \,p_2^3+ A_{050} \,p_1^5 \,+\,l.o.t.$ (case $Q_5$ in Ref.\cite{AW}) or $Y_5 =A_{032}\, p_1^3 \,p_2^2+ A_{005} \,p_2^5 \,+\,l.o.t.$. Similar comments apply to the cases $Q_7$, and $Q_9$ of Ref.\cite{AW}\ .

\subsection{Case $ N=6$}

\bigskip

For $N=6$, as previously mentioned the terms $A_{060}\,p_1^6,\, A_{006}\,p_2^6$ and $A_{033}\,p_1^3 \,p_2^3$ in (\ref{YN6de}) can be removed from $Y_6$ by means of the two trivial integrals $H$ and $X$. In fact, the term $A_{033}\,p_1^3 \,p_2^3$ leads to the square of a lower order superintegrable system. Here, we only consider the following case

\vspace{0.2cm}

$ \bullet \ Y_6\ = \ \{\,L_z^{2},\,p_1^2 \,p_2^2\,\} \ + \  \ldots$

\vspace{0.4cm}

This singlet is the second element in the family $II$ starting at $N=5$. The potential is given by
\begin{equation}\label{}
V(x,y)\ = \ \hbar^2\,[\,\mathcal{F}'(x;a,b) \ + \ \mathcal{F}'(y;\tilde a,\tilde{b})\,] \ ,
\end{equation}
where the function $\mathcal{F}=\mathcal{F}(z;a,b)$ satisfies the nonlinear ODE
\begin{equation}\label{N6A222}
\begin{aligned}
& \big[\,z^{2}\,{\mathcal{F}}^{(3)} \ + \ 2\,z\,{\mathcal{F}}^{(2)} \ - \  2\,{\mathcal{F}}'  \,\big] \ - \
 {\mathcal{F}}'\,\big[\,6\,z^{2}\,{\mathcal{F}}' \,+\,4\,z\,{\mathcal{F}} \,+\, a_1\,z\,+\, a_2\,z^2\,+\,\sigma\,z^4   \,\big]
  \ + \
\\ &
    {\mathcal{F}}\,\big[\,2\, {\mathcal{F}} \,+\,a_1 \,-\,2\,\sigma\,z^3 \,\big] \ + \ b_0 \ + \  b_2\,z^2 \ + \ b_3\,z^3 \ + \ b_4\,z^4 \ + \ \Lambda\,z^6  \ = \ 0 \ ,
\end{aligned}
\end{equation}
$\mathcal{F}^{(\ell)} \equiv \frac{d^\ell}{dz^\ell}\,\mathcal{F}(z)$, here the $a_k$, $b_k$, $\sigma$ and $\Lambda$ are real constants. The equation (\ref{N6A222}) passes the Painlev\'e test for any value of these parameters, the resonances occur at $r=1,6$. The solution of (\ref{N6A222}) can be expressed in terms of the fifth Painlev\'e transcendent function $P_5$ (see eqs. (42)-(45) in Ref.\cite{MSW}).

\vspace{0.4cm}

In the particular case when all the real constants in (\ref{N6A222}) are identically zero we obtain the solutions:

\begin{align}\label{}
V_1(x) \ = \ \frac{\hbar ^2}{2}\bigg(\sqrt{\alpha }\, P_3'\,+\,\frac{3}{4} \alpha \, P_3^2\,+\,\frac{\delta }{4\,P_3^2}\,+\,\frac{\beta \, P_3}{2\,x}\,+\,\frac{\gamma}{2\, P_3\,x}\,-\,\frac{ P_3'}{2\,x \,P_3}\,+\,\frac{P_3'^2}{4 \,P_3^2}\bigg) \ ,
\end{align}

\begin{align}\label{}
V_2(y) \ = \ \frac{\hbar ^2}{2}\bigg(\sqrt{ \alpha }\, P_3'\,+\,\frac{3}{4}  \alpha \, P_3^2\,+\,\frac{ \delta }{4\,P_3^2}\,+\,\frac{ \beta \, P_3}{2\,y}\,+\,\frac{\gamma}{2\, P_3\,y}\,-\,\frac{ P_3'}{2\,y \,P_3}\,+\,\frac{P_3'^2}{4 \,P_3^2}\bigg) \ ,
\end{align}

where $P_3=P_3(u)$ satisfies the third Painlev\'e equation
$$P_3''\ = \ \frac{P_3'^2}{P_3}\ -\ \frac{P_3'}{u}\ + \ \alpha \,P_3^3\ + \ \frac{\beta P_3^2+\gamma}{u}\ + \ \frac{\delta}{P_3}\ .$$

\bigskip

\subsection{Case $ N=7$}

\bigskip

\vspace{0.3cm}

$ \bullet \ Y_7^{(a)} \ = \ \{\,L_z^{3},\,p_1^2 \,p_2^2\,\} \ + \  \ldots$

\vspace{0.4cm}

This singlet is the third element in the family $II$ that starts at $N=5$. The potential is given by
\begin{equation}\label{}
V(x,y)\ = \ \hbar^2\,[\,\mathcal{F}'(x;a,b) \ + \ \mathcal{F}'(y;\tilde a,\tilde{b})\,] \ ,
\end{equation}
where the function $\mathcal{F}=\mathcal{F}(z;a,b)$ satisfies again a nonlinear ODE of the form
\begin{equation}\label{N7A322}
\begin{aligned}
& \big[\,z^{3}\,{\mathcal{F}}^{(3)} \ + \ 4\,z^2\,{\mathcal{F}}^{(2)} \ - \  8\,{\mathcal{F}}  \,\big] \ - \
 {\mathcal{F}}'\,\big[\,6\,z^{3}\,{\mathcal{F}}' \,+\,8\,z^2\,{\mathcal{F}} \,+\, a_1\,z\,+\, a_2\,z^2\,+\,\sigma\,z^5   \,\big]
  \ + \
\\ &
    {\mathcal{F}}\,\big[\,4\,z\, {\mathcal{F}} \,+\,2\,a_1\,+\, a_2\,z  \,-\,2\,\sigma\,z^4 \,\big] \,+\, b_0 \,+\, b_1\,z \,+\, b_3\,z^3  \,+\, b_4\,z^4   \,+\, b_5\,z^5  \ + \ \Lambda\,z^7  \ = \ 0 \ ,
\end{aligned}
\end{equation}
(cf.(\ref{N6A222})) $\mathcal{F}^{(\ell)} \equiv \frac{d^\ell}{dz^\ell}\,\mathcal{F}(z)$, here the $a_k$, $b_k$, $\sigma$ and $\Lambda$ are real constants. Again, the equation (\ref{N7A322}) passes the Painlev\'e test for any value of the parameters, the resonances occurring at $r=1,6$.

\vspace{0.6cm}

$\bullet \ Y_7^{(b)} \ = \ \cos \gamma\, p_1^7 \ + \ \sin \gamma\,p_2^7 \ + \  \ldots$

\vspace{0.4cm}

$0 < \gamma \neq \frac{\pi}{2}< \pi$. In this case, the function $V_1$ satisfies

\begin{align}
\label{}
\begin{aligned}
	& \cos \gamma\,\bigg[\hbar^6\, V_1{}^{(6)}  \ - \ 4\,\hbar^4\,V_1{}^{(4)} \left(7  \,V_1+c_1\right) \ + \ 40\, \hbar ^2 \,\left(V_1'\right){}^2 \left(7\,  V_1 \,+\,c_1\right) \\ & \   + \ 8 \,\hbar^2\, V_1'' \,\left(35 \, V_1^2+10\, c_1\, V_1+2\, c_2\right)
\ - \ 56\, \hbar^4\, V_1{}^{(3)} \, V_1' \ - \ 42 \, \hbar ^4\, \left(V_1''\right){}^2 \\ &
\ - \ 280\,  V_1^4 \ - \ 160\, c_1\, V_1^3\ - \ 96\, c_2\, V_1^2\ - \ 64\, c_3\, V_1\bigg] \ = \ \lambda \, x  \ ,
\end{aligned}
\end{align}

where the $c$'s and $\lambda$ are arbitrary constants. Similarly, the function $V_2(y)$ obeys

\begin{align}
\begin{aligned}
\label{}
	& \sin \gamma\,\bigg[\hbar^6\, V_2{}^{(6)}  \ - \ 4\,\hbar^4\,V_2{}^{(4)} \left(7\,V_2+b_1\right) \ + \ 40\, \hbar ^2 \,\left(V_2'\right){}^2 \left(7\, V_2 \,+\,b_1\right) \\ & \   + \ 8 \,\hbar^2\, V_2'' \,\left(35 \, V_2^2+10\, b_1\, V_2+2\, b_2\right)
\ - \ 56\, \hbar^4\, V_2{}^{(3)} \, V_2' \ - \ 42 \, \hbar ^4\, \left(V_2''\right){}^2 \\ &
\ - \ 280\,  V_2^4 \ - \ 160\, b_1\, V_2^3\ - \ 96\, b_2\, V_2^2\ - \ 64\, b_3\, V_2\ \,\bigg] \ = \ -\lambda \, y \ ,
\end{aligned}
\end{align}

in agreement with eq. $(3.40)$ in Ref.\cite{GungorKNN:2014}. Also, it is conjectured that the above two equations define a new transcendent. They pass the Painlev\'e test ($\hbar \neq 0$), the resonances occur at $r=2,4,5,7,10$.

\subsection{Case $N=8$}

\bigskip

\vspace{0.3cm}

$ \bullet \ Y_8 \ = \ \{\,L_z^{4},\,p_1^2 \,p_2^2\,\} \ + \  \ldots$

\vspace{0.4cm}

In this case, the potential is given by
\begin{equation}\label{}
V(x,y)\ = \  \hbar^2\,[\,\mathcal{F}'(x;a,b) \ + \ \mathcal{F}'(y;\tilde a,\tilde{b})\,]     \ ,
\end{equation}
where the function $\mathcal{F}=\mathcal{F}(z;a,b)$ satisfies the third-order nonlinear ODE
\begin{equation}\label{}
\begin{aligned}
& \bigg[\,z^4\,{\mathcal{F}}^{(3)} \ + \ 6\,z^3\,{\mathcal{F}}^{(2)} \ + \  6\,z^2\,{\mathcal{F}}' \ - \  24\,z\,{\mathcal{F}} \,\bigg] \ - \
 {\mathcal{F}}'\,(\,6\,z^4\,{\mathcal{F}}' \,+\,12\,z^3\,{\mathcal{F}} \,+\, z(a_1\,z+a_2) + \sigma\,z^6  \,)
\\ &
  \ + \   {\mathcal{F}}\,(\,6\,z^2\, {\mathcal{F}} \,+\,2\,z\,a_1 \,+\, 3\,a_2 \,-\,2\,\sigma\,z^5\,) \,+\, b_0 \,+\, b_1\,z \,+\, b_2\,z^2  \,+\, b_4\,z^4   \,+\, b_5\,z^5    \  + \ \Lambda\,z^8    \ = \ 0 \ .
\end{aligned}
\end{equation}

The above equation also passes the Painlev\'e test, the resonances occur at $r=1,6$.

\subsection{Case $ N=9$}

\bigskip

\vspace{0.3cm}

$ \bullet \ Y_9^{(a)} \ = \ \{\,L_z^{5},\,p_1^2 \,p_2^2\,\} \ + \  \ldots$

\vspace{0.4cm}

In this case, the potential reads
\begin{equation}\label{}
V(x,y)\ = \  \hbar^2\,[\,\mathcal{F}'(x;a,b) \ + \ \mathcal{F}'(y;\tilde a,\tilde{b})\,]     \ ,
\end{equation}
where the function $\mathcal{F}=\mathcal{F}(z;a,b)$ satisfies the third-order nonlinear ODE
\begin{equation}\label{N9Equ}
\begin{aligned}
& \bigg[\,z^5\,{\mathcal{F}}^{(3)} \ + \ 8\,z^4\,{\mathcal{F}}^{(2)} \ + \  16\,z^3\,{\mathcal{F}}' \ - \  48\,z\,{\mathcal{F}} \,\bigg] \, - \,
 {\mathcal{F}}'\,(\,6\,z^5\,{\mathcal{F}}' \,+\,16\,z^4\,{\mathcal{F}} \,+\, a_1\,z \,+\, a_2\,z^2\,+\,a_3\,z^3  + \sigma\,z^7  \,)
\\ &
  \ + \   {\mathcal{F}}\,(\,8\,z^3\, {\mathcal{F}} \,+\,4\,a_1 \,+\, 3\,a_2\,z\,+\,2\,a_3\,z^2 \,-\,2\,\sigma\,z^6\,) \,+\, b_0 \,+\, b_1\,z \,+\, b_2\,z^2  \,+\, b_3\,z^3
 \\ &
    \ + \  b_5\,z^5 \ + \  b_6\,z^6    \ + \ \Lambda\,z^9    \ = \ 0 \ .
\end{aligned}
\end{equation}

Not surprisingly, the equation (\ref{N9Equ}) passes the Painlev\'e test as well.

\vspace{0.3cm}

$ \bullet \ Y_9^{(b)} \ = \  \cos \gamma\, p_1^9 \ + \ \sin \gamma\,p_2^9  \ + \  \ldots$

\vspace{0.4cm}

The above integral belongs to the infinite family $I$. The corresponding potential can be expressed as

\begin{equation}\label{}
V(x,y)\ = \ U\bigg(x;\frac{\lambda}{\cos \gamma},c\bigg) \ + \ U\bigg(y;-\frac{\lambda}{\sin \gamma},\tilde c\bigg) \ ,
\end{equation}

($0< \gamma \neq \frac{\pi}{2}< \pi$) where the function $U=U(z;b,c)$ satisfies the eighth-order nonlinear ODE
\begin{equation}\label{B9I}
\begin{aligned}
& \hbar ^8\,U^{(8)} \ - \ \hbar^6\, \bigg[\,4 \,U^{(6)} \,\left(9 \,U \ + \ c_1\right)\ + \ 138\, \left(U^{(3)}\right)^2\ + \ 108\, U^{(5)}\, U'\ +\ 228\, U^{(4)} \,U''\,\bigg] \ + \
\\ &
\hbar^4\, \bigg[\,112\, U^{(4)}\, U \,c_1 \ + \ 16\, U^{(4)} \,c_2 \ + \ 168\, \left(9 \,U+c_1\right) \left(U''\right)^2\,+\,504\, U^{(4)} \,U^2 \ +\
\\ &
224\, U^{(3)} \left(9 \,U+c_1\right) U'\ +\ 1848\, \left(U'\right)^2\, U''\,\bigg] \ + \ \hbar ^2\, \bigg[\,-1120\,c_1  \,U^2\, U'' \ - \ 320\,c_2\, U \, U''\ - \ 64\, c_3\, U''\ - \
\\ &
3360\, U^3\, U''\ - \ 80\, \left(\,63\, U^2 \,+\,14\, U\, c_1 \,+\,2\, c_2\,\right)\, \left(U'\right)^2\,\bigg] \ + \ 2016 \,U^5 \ + \ 1120\, U^4\, c_1 \ + \
\\ &
640\, U^3\, c_2 \ + \ 384\, U^2\, c_3 \ + \ 256\, U\, c_4 \ = \ b\,z
    \nonumber \ ,
\end{aligned}
\end{equation}
here the $c$'s and $b$ are arbitrary constants. Independently of these constants, the equation (\ref{B9I}) passes the Painlev\'e test ($\hbar \neq 0$), the resonances occur at $r=2,4,5,6,7,9,12$.

\subsection{Case $ N=10$}

\vspace{0.2cm}

$ \bullet \ Y_{10}\ = \ \{\,L_z^{6},\,p_1^2 \,p_2^2\,\} \ + \  \ldots$

\vspace{0.4cm}

Finally, the potential is of the form

\begin{equation}\label{}
V(x,y) \ = \  \hbar^2\,[\,  \mathcal{F}'(x;a,b) \ + \ \mathcal{F}'(y;\tilde{a},\tilde{b})\,]     \ ,
\end{equation}
where the function $ \mathcal{F} =\mathcal{F}(z;a,b)$ satisfies the third-order nonlinear ODE
\begin{equation}\label{E10P}
\begin{aligned}
& \bigg[\,z^6\,{\mathcal{F}}^{(3)} \ + \ 10\,z^5\,{\mathcal{F}}^{(2)} \ + \  30\,z^4\,{\mathcal{F}}' \ - \  80\,z^3\,{\mathcal{F}} \,\bigg] \ - \
\\ &
 {\mathcal{F}}'\,(\,6\,z^6\,{\mathcal{F}}' \ + \ 20\,z^5\,{\mathcal{F}} \ + \ a_3\,z^3 \ +  \  a_2\,z^2 \ + \ a_1\,z \  + \ \sigma\,z^8  \,)   \ + \
\\ &
{\mathcal{F}}\,(\,10\,z^4\, {\mathcal{F}} \,+\,3\,z^2\,a_3 \,+\, 4\,z\,a_2 \,+\,5\,a_1\,-\, 2\,\sigma\,z^7 \,)\ + \ P(z,b)  \ + \ \Lambda\,z^{10}     \ = \ 0 \ .
\end{aligned}
\end{equation}
where $P(z,b)$ is a certain ninth-degree polynomial in $z$. The equation (\ref{E10P}) passes the Painlev\'e test, the resonances occur at $r=1,6$.

\bigskip

\section{CONCLUSIONS}
\label{Conclusions}

Let us sum up the main results reported in this paper.

\begin{enumerate}
  \item The linear compatibility condition necessary for the existence of an $N$th order integral of motion $Y$ is in general a PDE of order $N$ in two variables. For separable potentials (\ref{Hcart}) it reduces to a system of $4$ linear ODEs of order $(N+2)$, two for $V_1(x)$ and two for $V_2(y)$ (see Theorem 1 in Section \ref{LCC NLCC}). The equations for $V_1$ and $V_2$ are related by a permutation of $x$ and $y$ and the corresponding permutation $A_{N-m-n,m,n} \Leftrightarrow A_{N-m-n,n,m}$.
  \item We have used the above mentioned linear compatibility conditions to define standard, singly exotic a doubly exotic potentials for arbitrary $N\geq 3$. We concentrated on doubly exotic ones for which all 4 linear compatibility conditions for the potential functions $V_1(x)$ and $V_2(y)$ are satisfied trivially. Any further linear equations for the potential functions that may arise in the solution of the complete set of determining equations must also be forced to be trivial (by a suitable choice of the constants in these equations).
  \item The doubly exotic potentials must satisfy further compatibility conditions, these however are nonlinear. In Section \ref{coefj4}, we show how to obtain and successively solve a “well” of such nonlinear compatibility conditions separately for $V_1$ and $V_2$. The quantum doubly exotic potentials are proportional to $\hbar^2$. Thus, they differ from their classical counterpart.
  \item In Section \ref{trivint} we show that two types of trivial integrals of order $N$ exist for every integrable or superintegrable system. The first type is a consequence of separability alone (polynomials in $H$ and $X$) and do not lead to a superintegrable system. The second type is a consequence of superintegrability at order $M< N$ and does not lead to new superintegrable potentials.
  \item In Section \ref{partcases} we restrict to integrals $Y$ that respect the $x,y$ permutation symmetry mentioned above. Their leading terms have the form (\ref{PN1T}) or (\ref{PN2T}). Accordingly, we obtained two infinite families of such superintegrable systems.
  \item In Section \ref{DoublyExoticlow} we consider special cases, namely $N=3,4,\ldots,10$. From these examples we see that in each case we obtain one nonlinear ODE for $V_1(x)$ and one for $V_2(y)$. These equations always successfully pass the Painlev\'e test and in many cases have been shown to actually have the Painlev\'e property. This leads us to the main result of this article in the form of two conjectures.
\end{enumerate}

\begin{conjecture}
For an $N$th-order polynomial integral
\begin{equation}\label{}
Y_{N,DE}^{(I)}\ = \ \cos \gamma\,p_1^{N}\ + \ \sin \gamma \,p_2^{N} \ + \ \text{(lower order terms)}\ ,  \qquad \text{for odd} \ N\geq 3  \ ,
\end{equation}
$0 < \gamma \neq \frac{\pi}{2}< \pi$, an infinite family of doubly exotic quantum superintegrable potentials occur. The potential is given by
\begin{equation}\label{}
V(x,y)\ = \ \hbar^2\,\big[\,U(x;\lambda\,\cos^{-1} \gamma  ) \ + \ U(y;-\lambda\,\sin^{-1} \gamma)\,\big]\ ,
\end{equation}
where the function $U=U(z;b)$ satisfies the nonlinear ODE of order
$(N-1)$
\begin{equation}\label{F1N}
\begin{aligned}
& \,U^{(N-1)} \ + \ \sum_{k=1}^{\frac{N-3}{2}}\,\bigg[\, U^{(N-2k-1)}\,{\mathcal{P}}_k(U)
\\ &
 \ + \
 \sum_{r (i+1)+s( j+1)=N-2k-1}^{}{\big[U^{(i+1)}\big]}^r\,{\big[U^{(j+1)}\big]}^s\,{\mathcal{P}}_{k-1}(U)   \, \bigg]
\ + \  {\mathcal{P}}_{\frac{N+1}{2}}(U)   \ = \ b\,z
     \ ,
\end{aligned}
\end{equation}
$U^{(\ell)} \equiv \frac{d^\ell}{dz^\ell}\,U(z)$, here $b$ is a constant. The functions ${\mathcal{P}}_k={\mathcal{P}}_k(U;N),\,{\mathcal{P}}_{k-1}={\mathcal{P}}_{k-1}(U;N,i,j,s,r)$ and ${\mathcal{P}}_{\frac{N+1}{2}}={\mathcal{P}}_{\frac{N+1}{2}}(U;N)$ figuring in (\ref{F1N}) are polynomials in $U$, with constant coefficients, of order $k,k-1$ and $\frac{N+1}{2}$, respectively.
\end{conjecture}

\begin{conjecture}
For an $N$th-order polynomial integral
\begin{equation}\label{}
Y_{N,DE}^{(II)}\ = \ \{\,L^{(N-4)}_z,\,p_1^{2}\,p_2^{2}\,\} \ + \ \text{(lower order terms)}\ , \qquad \ N\geq 5 \ ,
\end{equation}
a second infinite family of doubly exotic quantum superintegrable potentials appears. In this case, the potential is given by
\begin{equation}\label{}
V(x,y)\ = \ \hbar^2\,[\,\mathcal{F}'(x;a) \ + \ \mathcal{F}'(y;b)\,] \ ,
\end{equation}
where the function $\mathcal{F}=\mathcal{F}(z;a)$ satisfies a nonlinear third order ODE of the form
\begin{equation}\label{F2N}
\begin{aligned}
& \big[\,z^{N-4}\,{\mathcal{F}}^{(3)} \ + \ 2(N-5)\,z^{N-5}\,{\mathcal{F}}^{(2)} \ + \  a_{N-6}\,z^{N-6}\,{\mathcal{F}}' \ + \  a_{N-7}\,z^{N-7}\,{\mathcal{F}} \,\big] \ - \
\\ &
 {\mathcal{F}}'\,\big[\,6\,z^{N-4}\,{\mathcal{F}}' \,+\,4\,(N-5)\,z^{N-5}\,{\mathcal{F}} \,+\, Q_{1}(z) \, +\, \sigma\,z^{N-2}    \,\big]
  \ + \
\\ &
    {\mathcal{F}}\,\big[\, 2\,(N-5)\,z^{N-6}\, {\mathcal{F}} \,+\,Q_{2}(z) -2\,\sigma\,z^{N-3} \,\big] \ + \ Q_{3}(z)  \ + \ \Lambda\,z^N       \ = \ 0 \ ,
\end{aligned}
\end{equation}
$\mathcal{F}^{(\ell)} \equiv \frac{d^\ell}{dz^\ell}\,\mathcal{F}(z)$, here the $a_k$'s are real constants and are identically zero for $k<0$. The functions $Q_{q}$'s figuring in (\ref{F2N}) are polynomials in $z$ of degree not larger than $(N-1)$. The parameters $\sigma$, $\Lambda$ are real constants.
\end{conjecture}

The conjectures have been confirmed for $N=3$ up to $N=10$. We conjecture it to be true for all $N$.

Several comments are in order.

Nonlinear ODEs may have special solutions that are also solutions of linear ODEs. For instance five of the six original Painlev\'e transcendents depend on between $1$ and $4$ complex parameters. For special values of these parameters they have so called classical solutions in terms of elementary functions, hypergeometric, cylindrical functions or other solutions of linear (see e.g. the book\cite{Laine}).

We do not claim that we have a complete classification of all doubly exotic quantum superintegrable systems of the type (\ref{Hcart}), (\ref{X}). Additional ones may appear for special vales of $N$ and even additional infinite families may exist.

In the case of doubly exotic classical systems, one can also obtain one nonlinear ODE for $V_1(x)$ and one for $V_2(y)$. Unlike the quantum case, these equations do not pass the Painlev\'e test and in many cases it can be shown that they actually reduce to pure algebraic equations. Work is in progress on the classical doubly exotic potentials and further properties of the quantum ones.

\section{ACKNOWLEDGMENTS}
The research of PW was partially supported by a research grant from NSERC of Canada.
AM is thankful to the Centre de Recherches Math\'{e}matiques, Universit\'{e} de Montr\'{e}al, for kind hospitality extended to
him where this work was initiated. RL is supported in part by CONACyT grant 237351 (Mexico).

{}


\bigskip



\begin{thebibliography}{}
\bibitem{ARS}
Ablowitz M, Ramani A and Segur H 1980 {A connection
between nonlinear evolution equations and ordinary differential
equations of P-type. I, II} {\it J. Math. Phys.} {\bf 21} 715 (1006)


\bibitem{AW}
Abouamal I and Winternitz P 2018 {Fifth-order superintergrable quantum system separating in Cartesian coordinates: Doubly exotic potentials} {\it J. Math. Phys.} {\bf 59} 022104
%
\bibitem{Atakishiyev}
Atakishiyev N M, Pogosyan G S, Wolf K B and Yakhno A 2019 {Spherical geometry, Zernike's separability, and interbasis expansion coefficients} {\it J. Math. Phys.} {\bf 60} 101701

\bibitem{Bargmann}
Bargmann V 1936 {Theory of the hydrogen atom} {\it Zeits. f. Physik} {\bf 99} 578

\bibitem{Benenti}
Benenti S, Chanu C and Rastelli G 2002 {Remarks on the connection between the additive separation of the Hamilton–Jacobi equation and the multiplicative separation of the Schrödinger equation. I, II. The completeness and Robertson conditions} {\it J. Math. Phys.} {\bf 43} 5183 (5223)


\bibitem{Boyer}
Boyer C P and Miller W Jr 1974 {A classification of second order raising operators for Hamiltonians in two variables} {\it J. Math. Phys.} {\bf 15} 1484-1489

\bibitem{Bureau1}
Bureau F J 1964 {Differential equations with fixed critical points} {\it Ann. Mat. Pura Appl.} {\bf LXIV} 229-364

\bibitem{Bureau2}
Bureau F J 1964 {Differential equations with fixed critical points} {\it Ann. Mat. Pura Appl.} {\bf LXVI} 1-116

\bibitem{Burchnall}
Burchnall J L and Chaudy T W 1928 {Commutative ordinary differential operators} {\it Proc. R. Soc. London} {\bf 118} 557-583; 1932 {\bf 134}  471-485

\bibitem{Campoamor}
Campoamor-Stursberg R, Cariñena J F and Ra\~{n}ada M F 2014 {Higher-order superintegrability of a Holt related potential} {\it J. Phys. A-Math. Gen.} {\bf 46} 435202

\bibitem{Chanu}
Chanu C M and Rastelli G 2017 {Extended Hamiltonians and shift, ladder functions and operators} {\it Annals of Physics} {\bf 386} 254-274

\bibitem{Chalykh}
Chalykh O A and Veselov A P 1990 {Commutative rings of partial differential operators and Lie algebras} {\it Commun. Math. Phys.} {\bf 126} 597-611

\bibitem{Chen}
Chen Z, Marquette I and Zhang Y Z 2019 {Superintegrable systems from block separation of variables and
unified derivation of their quadratic algebras} {\it Annals of Physics} {\bf 411} 167970

\bibitem{Conte}
Conte R and Musette M 2008 { The Painlev\'e Handbook} {\it Netherlands: Springer}

\bibitem{Conte:R99}
Conte R
\newblock  The Painlev\'e Approach to nonlinear Ordinary Differential
  Equations.
\newblock {\em The Painlev\'e property, one century later,} 77--180.
\newblock {\em Springer, New York}, 1999.

\bibitem{Cosgrove}
Cosgrove C M 2000 {Higher-order Painlev\'e equations in the polynomial class I. Bureau symbol B2} {\it Stud. Appl. Math.} {\bf 104} 1-65


\bibitem{Cosgrove:SD}
Cosgrove C M and Scoufis G
\newblock Painlev\'e classification of a class of differential equations of the
  second order and second degree.
\newblock {\em Stud. Appl. Math}, 88(1):25--87, 1993.


\bibitem{DSD}
Daboul J, Slodowy P and Daboul C 1993 {The Hydrogen algebra as centerless twisted Kac-Moody
algebra} {\it Phys. Lett. B} {\bf 317} 321--8


\bibitem{Daskol}
Daskaloyannis C 1991 {Generalized deformed oscillator and nonlinear algebras} {\it J. Phys. A-Math. Gen.} {\bf 24} 789--94


\bibitem{Doebner}
Doebner H-D and Zhdanov R Z 1999 {The stationary KdV hierarchy and so$(2,1)$ as a spectrum generating algebra} {\it J. Math. Phys.} {\bf 40} 4995

\bibitem{Drach1}
Drach J 1935 {{S}ur l'integration logique des \'equations de la dynamique \`{a} deux variables: Forces constructives. Int\'egrales cubiques. Mouvements dans le plan} {\it Comptes Rendus Acad. Sci.} {\bf 200} 22

\bibitem{Eichler}
Eichler M 1968 {A new proof of the Baker-Campbell-Hausdorff formula} {\it J. Math. Soc. Japan} {\bf 20} 23-25

\bibitem{AMJVPW2015}
Escobar-Ruiz A M, L\'opez Vieyra  J C and Winternitz P 2018 {Fourth order superintegrable systems separating in polar coordinates. I. Exotic Potentials} {\it J. Phys. A-Math. Theor.} {\bf 50} 495206


\bibitem{MTE:20182}
Escobar-Ruiz A M, Miller W Jr and Turbiner A V  2019 {Four-body problem in $d$-dimensional space: ground state, (quasi)-exact-solvability. IV } {\it Journal of Math Physics} {\bf 60} 062101

\bibitem{EWY}
Escobar-Ruiz A M, Winternitz P and Yurdu\c{s}en {\.{I}} 2018  {General $N{th}$-order superintegrable systems separating in polar coordinates} {\it J. Phys. A Math. Theor. \bf 51} 40LT01


\bibitem{Evans1991}
Evans N W 1991 {Group theory of the {S}morodinsky-{W}internitz system} {\it J. Math. Phys. \bf 32} 3369

\bibitem{Nikitin}
Fushchych W I and Nikitin A G 1997 {Higher symmetries and exact solutions of linear and nonlinear Schrödinger equation} {\it J. Math. Phys.} {\bf 38} 5944

\bibitem{Fock}
Fock V A 1935 {To the theory of the hydrogen atom} {\it Zeits. f. Physik} {\bf 98} 145

\bibitem{Fordy1}
Fordy A P 2007 {Quantum Super-Integrable Systems as Exactly Solvable Models} {\it SIGMA} {\bf 3} 025

\bibitem{Fordy2}
Fordy A P 2018 {Classical and Quantum Super-Integrability: From Lissajous Figures to Exact Solvability} {\it Physics of Atomic Nuclei} {\bf 81} 832–842

\bibitem{Fris:1965}
Fri{\v s} J, Mandrosov V, Smorodinsky Y A, Uhl{\'{\i}}{\v r} M and Winternitz P 1965 {{O}n higher symmetries in quantum mechanics} {\it Phys. Lett.} {\bf 16} 354--6

\bibitem{Gambier:Pain}
Gambier B
\newblock Sur les \'equations diff\'erentielles du second ordre et du premier
  degr\'e dont l'int\'egrale g\'en\'erale est \`a points critiques fixes.
\newblock {\em Acta Mathematica}, 33(1):1--55, 1910.


\bibitem{Genest}
Genest V X, Miki H, Vinet L and Guofu Y 2017 {A superintegrable discrete harmonic oscillator based on bivariate Charlier polynomials} {\it Physics of Atomic Nuclei} {\bf 80} 794-800

\bibitem{gW}
Gravel S 2004 {{H}amiltonians separable in Cartesian coordinates and third-order integrals of motion} {\it J. Math. Phys.} {\bf 45} 1003-1019

\bibitem{gWW}
Gravel S and Winternitz P 2002 {{S}uperintegrability with third-order integrals in quantum and classical mechanics} {\it J. Math. Phys.} {\bf 43} 5902-5912

\bibitem{GriTsi}
Grigoryev Yu A and Tsiganov A V 2018 {On superintegrable systems separable in Cartesian coordinates} {\it Phys. Lett. A} {\bf 382} 2092-2096

\bibitem{Laine}
Gromak V I, Laine L, and Shimomura S 2013 {Painlev\'e Differential Equations in the Complex Plane} {\it Degruyter Studies in Mathematics} Reprint 2013 ed. Edition

\bibitem{Gonera}
Gonera C and Gonera J 2020 {New superintegrable models on spaces of constant curvature} {\it Annals of Physics} {\bf 413} 168052

\bibitem{GungorKNN:2014}
G\"{u}ng\"{o}r F, Kuru {\c S}, Negro J and Nieto L M 2017 {{H}eisenberg-type higher order symmetries of superintegrable systems separable in cartesian coordinates} {\it Nonlinearity} {\bf 30} 1788-1808

\bibitem{Hietarinta1987}
Hietarinta J 1987 {Direct methods for the search of the second invariant} {\it Phys. Rep.} {\bf 147} 87-154

\bibitem{Hietarinta1998}
Hietarinta J 1998 {Pure quantum integrability} {\it Phys. Lett. A} {\bf 246} 97-104

\bibitem{HietarintaGrammaticos}
Hietarinta J and Grammaticos B 1989 {On the $\hbar^2$ correction terms in quantum integrability} {\it J. Phys. A-Math. Gen.} {\bf 22} 1315--22


\bibitem{Iliev}
Iliev P 2018 {Symmetry algebra for the generic superintegrable system on the sphere} {\it Journal of High Energy Physics} {\bf 2} 044


\bibitem{Ince:ode}
Ince E L,
\newblock Ordinary differential equations.
\newblock {\em Dover, New York}, 1956.

\bibitem{Jauch}
Jauch J and Hill E 1940 {On the problem of degeneracy in quantum mechanics} {\it Phys. Rev.} {\bf 57} 641--5

\bibitem{Millerebook}
Kalnins E G, Kress J M and Miller W Jr 2018 {\it Separation of variables and Superintegrability: The symmetry of solvable systems} (UK,  ISBN: 978-0-7503-1314-8: Instititute of Physics)

\bibitem{Milleretal}
Kalnins E G, Kress J M and Miller W Jr 2012 {Structure relations for the symmetry algebras of quantum superintegrable systems} {\it J. Phys.: Conf. Ser.} {\bf 343} 012075

\bibitem{Kress}
Kress J M and Schoebel K 2019 {An algebraic geometric classification of superintegrable systems in
the Euclidean plane} {\it Journal of pure and applied algebra} {\bf 223} 1728-1752


\bibitem{Latini}
Latini D 2019 {Universal chain structure of quadratic algebras for superintegrable
systems with coalgebra symmetry} {\it J. Phys. A-Math. Gen.} {\bf 52} 125202



\bibitem{Magnus}
Magnus W 1954 {On the exponential solutions of differential equations for a linear operator} {\it Comm. Pure Appl. Math.} {\bf 7} 649-673


\bibitem{IanMcubic}
Marquette I 2009 {Superintegrability with third order integrals of motion, cubic algebras,
and supersymmetric quantum mechanics. II. Painleve transcendent potentials} {\it J. Math. Phys.} {\bf 50} 095202

\bibitem{IanMladder}
Marquette I 2010 {Construction of classical superintegrable systems with higher order
integrals of motion from ladder operators} {\it J. Math. Phys.} {\bf 51} 072903


\bibitem{IanM}
Marquette I 2012 {Classical ladder operators, polynomial Poisson algebras, and classification of superintegrable systems} {\it J. Math. Phys.} {\bf 53} 012901

\bibitem{MSW}
Marquette I, Sajedi M and Winternitz P 2017 {Fourth order superintegrable systems separating in Cartesian coordinates I. Exotic quantum potentials} {\it J. Phys. A-Math. Theor.} {\bf 50} 315201

\bibitem{MSW2}
Marquette I, Sajedi M and Winternitz P 2019 {Two-dimensional superintegrable systems from operator algebras
in one dimension} {\it J. Phys. A-Math. Theor.} {\bf 52} 115202

\bibitem{MPPC}
Marquette I and Winternitz P 2019 {Higher Order Quantum Superintegrability: a new "Painlev\'e conjecture"} {\it Integrability, Supersymmetry and Coherent States. CRM Series in Mathematical Physics. Springer, Cham}

\bibitem{MillerPostWinternitz:2013}
Miller W Jr, Post S and Winternitz P 2013 {Classical and quantum superintegrability with applications} {\it J. Phys. A-Math. Gen.} {\bf 46} 423001

\bibitem{MTE:2018}
Miller W Jr, Turbiner A V and Escobar-Ruiz M A 2018 {The quantum $n$-body problem in dimension $d\geq n-1$: ground state } {\it J. Phys. A: Math. Theor.} {\bf 51} 205201


\bibitem{Nikitin2}
Nikitin A G 2015 {Superintegrable and shape invariant systems with position dependent mass} {\it J. Phys. A: Math. Theor.} {\bf 48} 335201


\bibitem{Painleve:Pain}
Painlev\'e P
\newblock Sur les \'equations diff\'erentielles du second ordre et d'ordre
  sup\'erieur dont l'int\'egrale g\'en\'erale est uniforme.
\newblock {\em Acta Mathematica}, 25(1):1--85, 1902.

\bibitem{Pogosyan}
Pogosyan G S, Salto-Alegre C, Wolf K B and Yakhno A 2017 {Quantum superintegrable Zernike system} {\it J. Math. Phys.} {\bf 58} 072101


\bibitem{PostWinternitz:2015}
Post S and Winternitz P 2015 {General $N^{th}$-order integrals of motion in the Euclidean plane} {\it J. Phys. A-Math. Theor.} {\bf 48} 405201

\bibitem{Ranada}
Ra{\~n}ada M F 1997 {Superintegrable $n=2$ systems, quadratic constants of motion, and potentials of Drach} {\it J. Math. Phys.} {\bf 38} 4165--78

\bibitem{Shmavonyan}
Shmavonyan H 2019 {$\mathbb{C}^N$-Smorodinsky-Winternitz system in a constant magnetic field} {\it Phys. Lett. A} {\bf 383} 1223-1228


\bibitem{TempTW}
Tempesta P, Turbiner A V and Winternitz P 2001 {Exact solvability of superintegrable systems} {\it J. Math. Phys.} {\bf 42} 419--436


\bibitem{Tremblay}
Tremblay F, Turbiner A V and Winternitz P 2009 {An infinite family of solvable and integrable quantum systems on a
  plane} {\it J. Phys. A-Math. Theor.} {\bf 42} 242001

\bibitem{TPW2010}
Tremblay F and Winternitz P 2010 {Third-order superintegrable systems separating in polar coordinates} {\it J. Phys. A-Math. Theor.} {\bf 43} 175206

\bibitem{Tsiganov:2000}
Tsiganov A V 2000 {{T}he {D}rach superintegrable systems} {\it J. Phys. A-Math. Gen.} {\bf 33}  7407

\bibitem{TME3B}
Turbiner A V, Miller W Jr and Escobar-Ruiz A M 2019 {Superintegrable three-body problems} {\it preprint arXiv:1912.05726} [math-ph] 10 pages

\bibitem{Tyc}
Tyc T and Danner A J 2017 {Absolute optical instruments, classical superintegrability, and
separability of the Hamilton-Jacobi equation} {\it Phys. Rev. A} {\bf 96} 053838


\bibitem{Veselov}
Veselov A P and Shabat A B 1993 {Dressing chains and the spectral theory of the Schr\"{o}dinger operator} {\it Functional Analysis and its Applications } {\bf 27} 81-96


\bibitem{Weigert}
Weigert S 1992 {The problem of quantum integrability} {\it Physica D} {\bf 56} 107-119



\end{thebibliography}
\end{document}